\documentclass[useAMS,usenatbib]{mn2e}
\usepackage[pass,letterpaper]{geometry}
\usepackage{amssymb}
\usepackage{amsmath}
\usepackage{bbm}
\usepackage{txfonts}
\usepackage{psfrag}
\usepackage{graphicx}
\usepackage{natbib}
\usepackage{if then}
\usepackage{longtable}
\usepackage{color}
\usepackage{tikz}
\usetikzlibrary{shapes,arrows}
\usepackage{verbatim}
\usepackage{bm}
\usepackage{multirow}

\def\del#1{{}}
\def\mnras{MNRAS}

\newcommand{\apj}{ApJ}

\newcommand{\aap}{A{\&}A}

\newcommand{\prd}{Phys. Rev. D}

\newcommand{\apjs}{ApJS}

\newcommand{\aj}{AJ}

\newcommand{\jcap}{JCAP}
\newcommand{\nat}{Nature}

\newcommand{\vect}[1]{\bm{#1}}

\newcommand{\borg}{{\tt BORG}}
\newcommand{\ares}{{\tt ARES}}
\newcommand{\ktmpp}{K_\text{2M++}}
\newcommand{\Mktmpp}{{M^K_\text{2M++}}}
\newcommand{\Mpch}{$h^{-1}$~Mpc~}
\newcommand{\hMpc}{$h$~Mpc$^{-1}$}
\newcommand{\LCDM}{$\Lambda$CDM}

\renewcommand{\deg}{\ensuremath{{}^\circ}}

\newcommand{\tmpp}{{2M++}}
\newcommand{\diva}{{\sc diva}}

\title{Unmasking the Masked Universe: the 2M++ catalogue through Bayesian eyes}
\author[G. Lavaux \& J. Jasche]
       {Guilhem Lavaux $^{1}$ \& Jens Jasche $^{1,2}$ \\
$^{1}$ CNRS \& Sorbonne Universit\'{e}s, UPMC Univ Paris 06, UMR7095, Institut d'Astrophysique de Paris, F-75014, Paris, France \\
$^{2}$ Excellence Cluster Universe, Technische Universit\"at M\"unchen, Boltzmannstrasse 2, 85748 Garching, Germany\\
}
\begin{document}

\date{Accepted 2015 October 25. 
      Received 2015 September 16;
      in original form 2015 February 16}

\pagerange{\pageref{firstpage}--\pageref{lastpage}} \pubyear{2014}

\bibliographystyle{mn2e}

\maketitle

\label{firstpage}

\begin{abstract}
This work describes a full Bayesian analysis of the Nearby Universe as traced by galaxies of the 2M++ survey.  
The analysis is run in two sequential steps. The first step self-consistently derives the luminosity dependent galaxy biases, the power-spectrum of matter fluctuations and matter density fields within a Gaussian statistic approximation. The second step makes a detailed analysis of the three dimensional Large Scale Structures, assuming a fixed bias model and a fixed cosmology. This second step allows for the reconstruction of both the final density field and the initial conditions at $z=1000$ assuming a fixed bias model. From these, we derive fields that self-consistently extrapolate the observed large scale structures. We give two examples of these extrapolation and their utility for the detection of structures: the visibility of the Sloan Great Wall, and the detection and characterization of the Local Void using {\sc DIVA}, a Lagrangian based technique to classify structures. 
\end{abstract}

\begin{keywords}
methods: data analysis -- methods: statistical -- galaxies: statistics -- large-scale structure of Universe
\end{keywords}

\section{Introduction}
Over the last decades, the wealth of galaxy redshift catalogues has stupendously increased. Nowadays millions of galaxies with precision positioning on the sky and accurate redshifts are available and have to be handled and processed on a routinely basis. For example the Sloan Digital Sky Survey \citep[SDSS, e.g. ][]{YORK2000,SDSS7,SDSS10} provides millions of galaxy redshifts and the Six Degree Field Galaxy Redshift Survey \citep[6DFGRS][]{Jones09}, covering the southern sky, contains nearly 70~000 galaxies with accurate redshift measurements. While the amount of data has steadily increased, progress in the development of modern data analysis techniques has only been made in recent years. These advances are particularly crucial to interpret evermore complex data sets where time evolution of objects (e.g. star formation rate), non-linear dynamics (e.g. galaxy cluster formation), foreground subtraction as well as systematic selection effects become increasingly important. 

Inferring 3d density fields in a formal and rigorous Bayesian framework has several advantages. The first and foremost advantage is that all observational aspects are treated self-consistently yielding inferred 3d density fields that do not require any post-analysis correction. The second advantage is that the model yields more information on the density field than what is readily usable in catalogues. For example the tidal field created by visible large scale structures may trigger the collapse in other unobserved area of the Universe. This can raise the interesting possibility of predicting where structures (such as walls, filaments, clusters and voids) form. The actual presence of such inferred structures can then be tested via dedicated observations a posteriori.  Specifically this work focuses on developing a probabilistic structure predictor. We will concentrate on the void aspect in difficult unobserved regions like the Galactic plane. To characterize these voids we will make use of the previously presented {\sc DIVA} framework \citep{LW10}.

Particularly successful approaches to solving such ill-posed inverse problems rely on the Bayesian formulation of parameter inference. We define a forward data model that indicates how a continuous three-dimensional density field is transformed into a set of predicted observables which are then directly compared to data.  In our case the observable is the number density of galaxies in comoving space. Conversely, given the position of galaxies we may infer this density field provided it is decomposed on an adequate finite basis. In this context, the data model should include everything that may happen between the density field to the detection of a galaxy by an observer, which includes for example photon detection, galaxy detection efficiency. The full problem cannot be solved in its entirety but for sufficiently well constructed samples only basic selection criterion, such as flux limitation and overall redshift completeness, are important. 

Even in this optimistic context, the parameter inference problem is daunting: for typical inferences we need to treat on the order of $10^6 - 10^7$ highly degenerate parameters comprised typically of density per volume elements and power spectrum values. There exists a wealth of literature on the derivation of power spectra and correlation functions from noisy and incomplete data \citep[see e.g.][]{LS93,TEGMARK_2004,PERCIVAL2005}. However they never fully grasp the complexity of the posterior of a blind analysis of power spectra in data.  More recent developments, notably stimulated by the requirement of the Cosmic Microwave Background community \citep[see e.g.][]{Eriksen04,Jewell04,WANDELT2004}, have pushed the limits of density and power-spectrum reconstruction for galaxy redshift catalogue \citep{JASCHESPEC2010,JASCHE2010HADESMETHOD,JL14}. 

All the aforementioned techniques still require a good knowledge on how tracers have been selected.  To have the largest, deepest and cleanest galaxy redshift compilation we propose to use the \tmpp{} \citep{LH11} galaxy compilation. This survey offers a near full sky coverage at a magnitude $\ktmpp\leq 11.5$ and above 50\% coverage for $\ktmpp \leq 12.5$. Evolutionary effects of galaxies were corrected in average and the selection is done for a consistent population of galaxy. Finally, redshift completeness maps are provided for the two magnitude selections. 

The data application presented in this works builds upon our previously developed Bayesian data analysis algorithms \ares{} \citep[Algorithm for REconstruction and Sampling, ][]{JaschePspec2013} and \borg{} \citep[Bayesian Origin Reconstruction from Galaxies, ][]{JASCHEBORG2012}. Both these algorithms perform a Bayesian analysis of the 3d distribution of galaxies albeit with different assumptions on the noise and on the dynamics of the tracers. This work is structured as follows. In Section~\ref{sec:data}, we give a description of the \tmpp{} galaxy compilation which is the data that we are aiming at modelling. Then in Section~\ref{sec:methodology}, we present the pipeline and give a reminder on the working of the \ares{} and \borg{} models and algorithms. In Section~\ref{sec:Bayesian_inference}, we present the setup and the convergence tests of the Bayesian inference. In Section~\ref{sec:inference_results}, we analyse the results in the context of cosmography and structure classification. Finally, in Section~\ref{sec:Summary_an_Conclusion} we conclude.

\section{The 2M++ Survey}
\label{sec:data}
In this work we follow a similar procedure as described in \citet{JASCHE2010HADESDATA} and more recently in \citet{JLW15}, by applying the \borg{} algorithm to the 2M++ galaxy compilation \citep{LH11}. The 2M++ is a superset of the 2MASS Redshift Survey \citep[2MRS,][]{Huchra12}, with a greater depth and a higher sampling than the IRAS Point Source Catalogue Redshift Survey \citep[PSCZ,][]{Saunders00}. The photometry is based primarily on the Two-Micron-All-Sky-Survey (2MASS) Extended Source Catalogue
\citep[2MASS-XSC,][]{Skrutskie06}, an all-sky survey in the $J$, $H$ and $K_S$ bands. Redshifts in the $K_S$ band of the 2MASS Redshift Survey
(2MRS) are supplemented by those from the Sloan Digital Sky Survey Data Release Seven \citep[SDSS-DR7, ][]{SDSS7}, and the Six-Degree-Field Galaxy Redshift Survey Data Release Three \citep[6dFGRS,][]{Jones09}.
Data from SDSS were matched to that of 2MASS-XSC using the NYU-VAGC catalogue \citep{BLANTON2005}. As the
2M++ draws from multiple surveys, galaxy magnitudes from all sources were first recomputed by measuring the apparent magnitude in the $K_S$ band within a circular isophote at 20~mag~arcsec$^{−2}$ . Following a prescription described in \cite{LH11}, magnitudes were then corrected for Galactic extinction, cosmological surface brightness dimming and stellar evolution. After corrections the sample was limited to $\ktmpp \le 11.5$ in regions not covered by the 6dFGRS or the SDSS, and limited to $\ktmpp \le 12.5$ elsewhere. Other relevant corrections which were made to this
catalogue include accounting for incompleteness due to fibre-collisions in 6dF and SDSS, as well as treatment of the zone of avoidance (ZoA). Incompleteness due to fibre-collisions was treated by cloning redshifts of nearby galaxies
within each survey region as described in \cite{LH11}.

The treatment of the ZoA in the 2M++ will be ignored for this work as the Bayesian machinery naturally and self-consistently accounts for incomplete observations. The galactic plane will thus be simply obscured, the objects marked as cloned removed from the catalogue and the completeness set to zero in that region. The ZoA is defined in the 2M++ as the region delimited by $|b| \le 5^\circ$ for $l > 30^\circ$ and $l < 330^\circ$, and $|b| \le 10^\circ$ for $l \le 30^\circ$ or $l \ge 330^\circ$.

\begin{figure*}
	\begin{center}
		\includegraphics[width=.45\hsize]{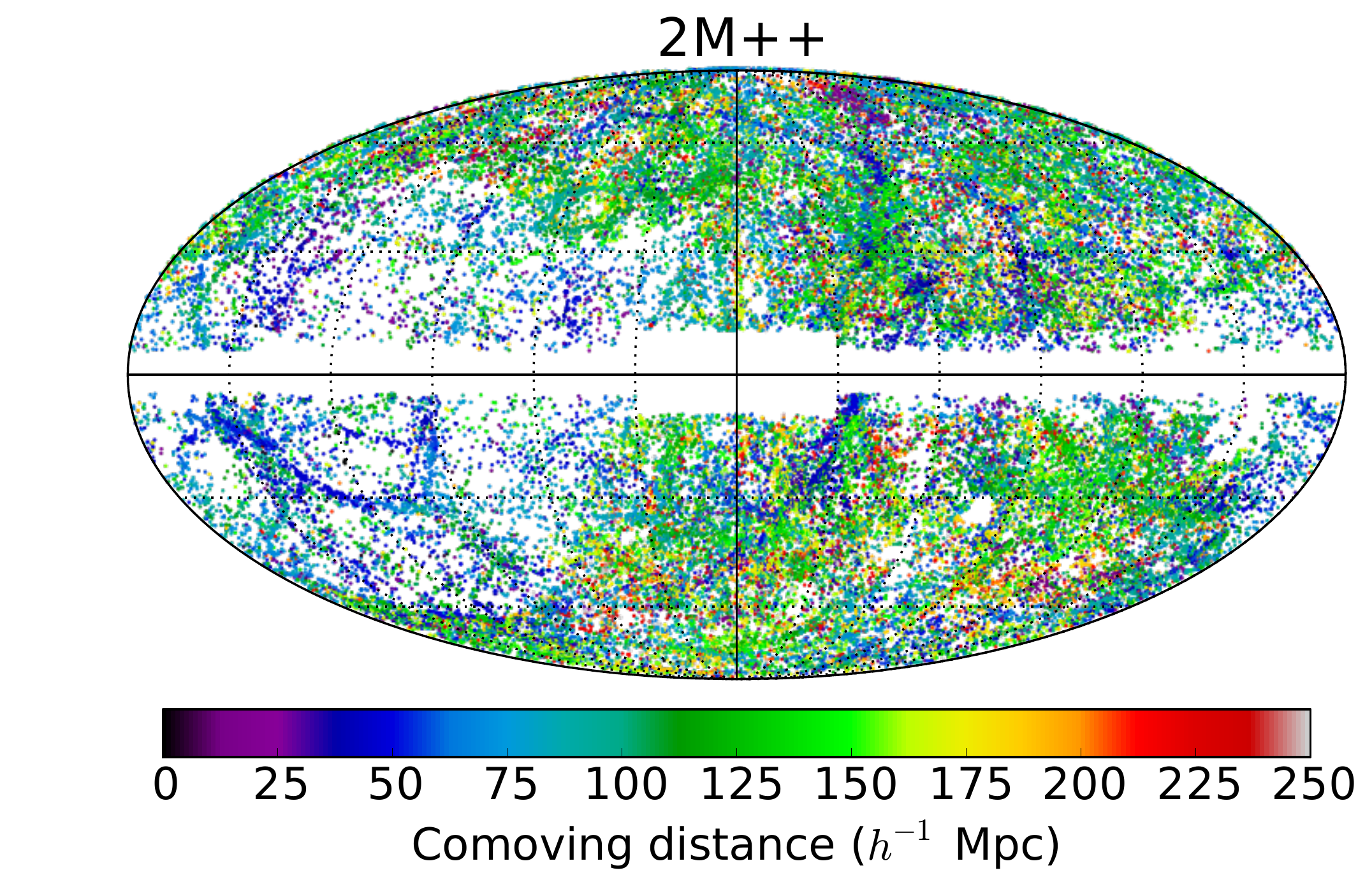}\\
		\includegraphics[width=.45\hsize]{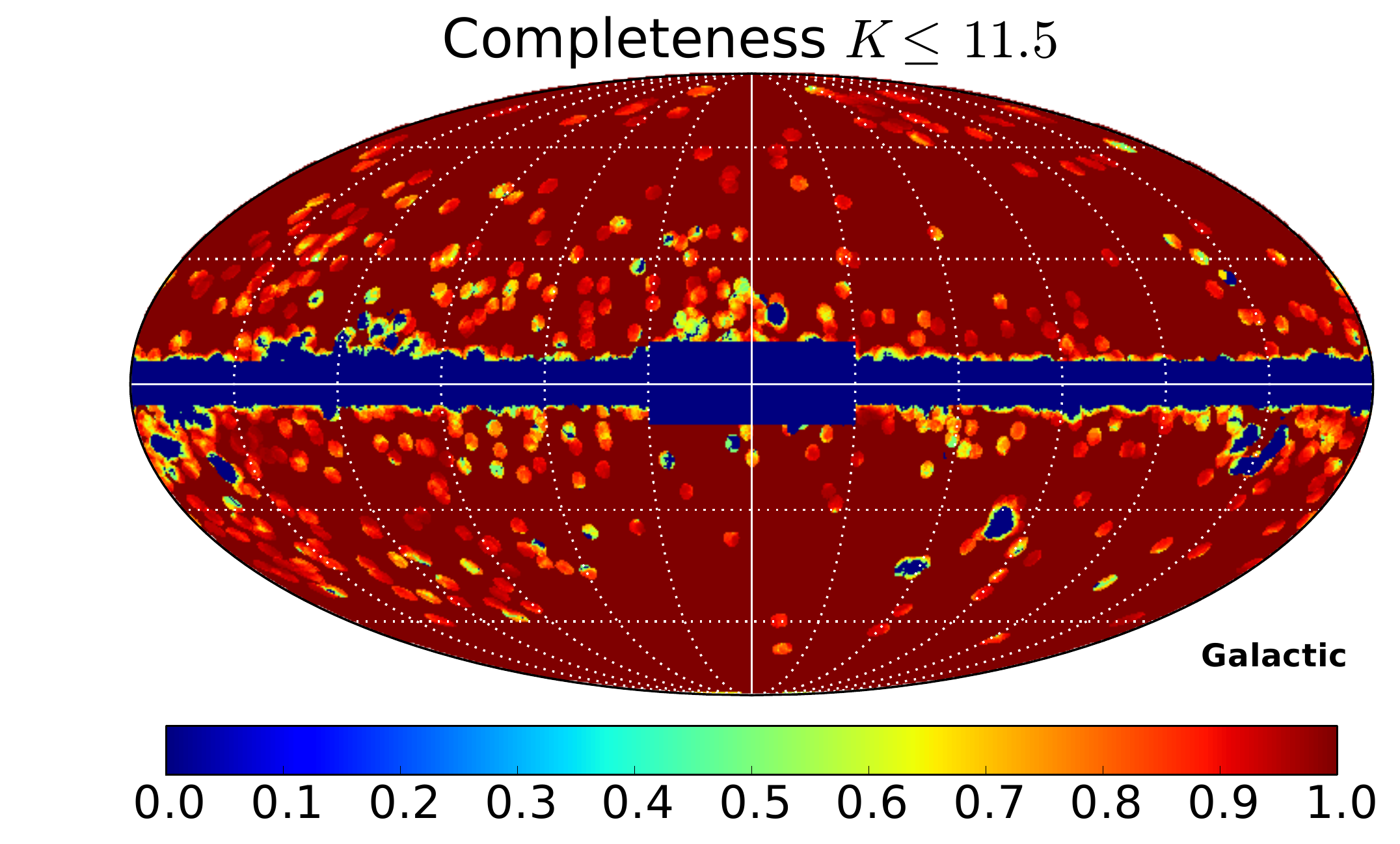}
		\includegraphics[width=.45\hsize]{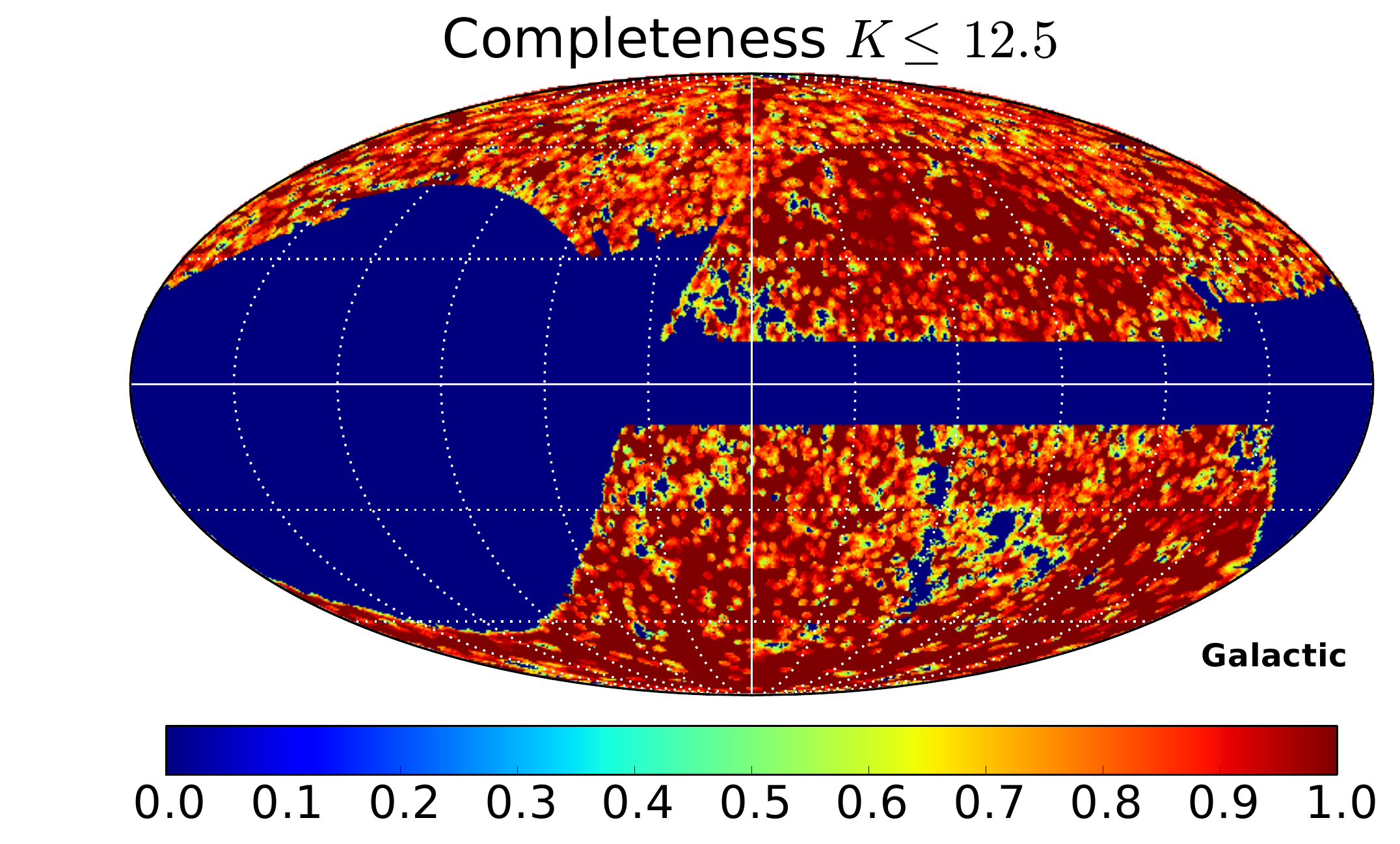}
	\end{center}
	\caption{\label{fig:2m++_data} We show here the 2M++ galaxy compilation. All plots uses a Galactic coordinate system. Top row: the 69081 galaxies that we used in the \ares{} and \borg{} analysis. Each galaxy is colour coded according to its apparent redshift. Bottom row: Redshift incompleteness mask for the two magnitude cuts $\ktmpp \le 11.5$ and $11.5 < \ktmpp \le 12.5$. Blue corresponds to zero completeness, which is equivalent in our scheme to be masked out.  }
\end{figure*}

The galaxy distribution on the sky and the corresponding selection at $\ktmpp \le 11.5$ and $11.5 < \ktmpp \le 12.5$ are given in Figure~\ref{fig:2m++_data}. The top row shows the data used in our analysis. The lower row show the redshift incompleteness, i.e. the number of acquired redshifts versus the number of targets, for the two apparent magnitude bins. We note that the galactic plane clearly stands out and that the incompleteness is evidently inhomogeneous and strongly structured. 

{In addition to the target magnitude incompleteness, and the redshift angular incompleteness, one may also worry about the dependence of the completeness with redshift. This is not a problem for the lower $\ktmpp \le 11.5$ which is essentially 100\% complete. We do not expect much effect in the fainter magnitude bins as the spectroscopic data come from SDSS and 6dFGRS which have both an homogeneous sampling and have fainter magnitude limits as the 2M++.}

We account for radial selection functions using a a standard luminosity function $\Phi(L)$ proposed by \cite{SCHECHTER1976}. Using this function we can deduce the expected number of galaxies in the absolute  magnitude range, observed within the apparent magnitude range of the sample at a given redshift. The $\alpha$ and $M^*$ parameters are given for the K$_S$-band in the line labeled "$|b|>10, K < 11.5$" of the table 2 of \cite{LH11}, i.e. $\alpha=-0.94$, $M^*=-23.28$. The target selection completeness of a voxel, indexed by $p$, is then
\begin{equation}
	c^t_p = \frac{\int_{\mathcal{V}_p} \text{d}^3 \vect{x} \int_{L_\text{app}(|\vect{x}|)}^{L_\text{max}} \Phi(L)\text{d}L}{V_p \int_{L_\text{min}}^{L_\text{max}} \Phi(L)\text{d}L},
\end{equation}
where $\mathcal{V}_p$ the co-moving coordinate set spanned by the voxel, and $V_p = \int_{\mathcal{V}_p} \text{d}^3 \vect{x}$.
The full completeness of the catalogue is derived from the product of $c^t$ and the map corresponding to the considered apparent magnitude cut given in the bottom row of the Figure~\ref{fig:2m++_data} after its extrusion in three dimensions.

Finally, we note that our analysis accounts for luminosity dependent galaxy biases by following the approach as described in \citet{JLW15}. In order to do so the galaxy sample is subdivided into 3 equidistant bins
in absolute $K$-band magnitude in the range $-25<\ktmpp <-21$. The galaxy sample is further splitted into two sub-sets depending on the apparent magnitude: if $\ktmpp \le 11.5$ it belongs to the sample one, otherwise, $11.5 < \ktmpp \le 12.5$ it belongs to the sample two. The bias in each of these bins is kept constant to greatly reduce the time complexity burden, at the cost of losing a full marginalization according to these parameters. The determination of these values is left to \ares{}. The mean density of tracers, and thus the Poisson noise amplitude, in each of these bins is sampled.

As will be described in more detail below, splitting the galaxy sample permits us to treat each of these sub-samples as an individual data set, with its respective selection effects, biases and noise levels.

\section{Methodology}
\label{sec:methodology}
In this section we give a brief introduction to the Bayesian inference framework \borg{} (Bayesian Origin Reconstruction from Galaxies).

\subsection{The \ares{} framework}

The \ares{} framework is a full Bayesian large scale structure inference method targeted at precision recovery 
of cosmological power-spectra from three dimensional galaxy redshift surveys. Specifically it performs joint inferences of three dimensional density fields, cosmological power spectra as well as luminosity dependent galaxy biases and corresponding noise levels for different galaxy populations in the survey \citep[][]{JASCHESPEC2010,JaschePspec2013}. 

The complete problem solved by \ares{} has many parameters. In the case of a single population, the data model implemented in \ares{} corresponds to the following:
\begin{equation}
	N_i = \bar{N} R_i (1 + b D_i \delta_i) + \epsilon_i,
\end{equation}
with $N_i$ the number of galaxies in the voxel $i$, $\bar{N}$ the mean density of the galaxy population, $R_i$ the overall linear response operator of the survey (i.e. the redshift and the target completeness), $b$ the population bias, $D_i$ the density growth factor in the voxel $i$, $\delta_i$ the linear density at a reference redshift in the voxel $i$ and $\epsilon_i$ a random instrumental noise. The noise is assumed to be Poissonian but approximated by a Gaussian distribution and neglecting the influence of the density fluctuations themselves. Thus we have
\begin{equation}
	\langle \epsilon_i \epsilon_j \rangle = \bar{N} R_i \delta^K_{i,j},
\end{equation}
with $\delta^K_{i,j} = 1$ is one for $i=j$ and zero otherwise. Finally, we add an isotropic Gaussian prior to $\delta_i$. All the details of the general model and the posterior formulation are given in \cite{JaschePspec2013}. {The linear bias model should be generally adequate to model the largest scale density fluctuations. In that regime, through Taylor expansion, all bias models are equivalent. However this is not the case at the smallest scales considered here ($\sim 2.3$\Mpch) though we think that we should not be strongly biased by this assumption. Effectively, we expect the signal-to-noise to ratio of the measurement of density modes to peak at intermediate scales ($\sim 10$\Mpch) and decreasing sharply both at small (for Poisson sampling reasons) and large (for selection reasons) scales. Thus the measured bias should actually represent the one at this typical scale. We finally note that the final confirmation that the bias model is not causing problems is the a posteriori confirmation that the recovered power spectrum is in agreement on large scales. 
}

{
To summarize the posterior from which we want to draw samples is
\begin{multline}
	\log \mathcal{P}(\delta_i, \bar{N}, b, P(k)| N_i) = C - \sum_i \frac{(\bar{N} R_i (1 + b D_i \delta_i) - N_i)^2}{2 \bar{N} R_i}  \\
	+ M \log \bar{N} - \sum_k \left(\frac{|\hat{\delta}_k|^2}{2 P_k} - \frac{1}{2} \log P_k\right),
\end{multline}
with $M$ the number of free voxels with non vanishing selection $R_i$, and $P_k$ the discrete powerspectrum of the density field. Such a posterior probability is too complex to analyse directly.}
In order to provide full Bayesian uncertainty quantification the algorithm explores the joint posterior distribution of all these quantities via an efficient implementation of high dimensional Markov Chain Monte Carlo methods in a block sampling scheme. In particular the sampling consists in generating from a Wiener posterior random realizations of three dimensional density fields {$\{\delta_i\}$ constrained by data $\{N_i\}$. Following each generation, we produce conditioned random realizations of the power-spectrum $\{ P_k \}$, galaxy biases $\{ b_q \}$ and noise levels $\{ \bar{N}_q \}$ through several sampling steps. }
Iteration of these sampling steps correctly yields random realizations from the joint posterior distribution.
In this fashion the \ares{} algorithm accounts for all joint and correlated uncertainties between all inferred quantities and allows for accurate inferences from galaxy surveys with non-trivial survey geometries.
Classes of galaxies with different biases are treated as separate sub samples, allowing even for combined analyses of more than one galaxy survey. 

This methodology has also been demonstrated to correctly treat anti-correlations between bias amplitudes and power spectrum, which are not taken into account in traditional approaches to power spectrum estimation, a 20 percent effect across large ranges in Fourier space \citep[][]{JaschePspec2013}.
In this work we use an upgraded version of the \ares{} which employs the messenger method discussed in \citet{2013A&A...549A.111E}. This particular implementation of the Wiener posterior sampling has been demonstrated to improve upon the statistical efficiency of previous implementations \citep[][]{JL14}.  
In this work we use the \ares{} algorithm to infer and calibrate luminosity dependent galaxy biases for the \tmpp{} galaxy survey.

\subsection{The \borg{} algorithm}

In addition to \ares{}, this work also capitalizes on the \borg{} \citep[Bayesian Origin Reconstruction from Galaxies][]{JASCHEBORG2012} algorithm to perform a chrono-cosmographical analysis of the 2M++ galaxy survey. The \borg{} algorithm is a fully probabilistic inference machinery aiming at the analysis of linear and mildly-non-linear matter density fields in galaxy observations. The algorithm incorporates a physical model for gravitational structure formation, which translates the traditional task of reconstructing the 3d density field into the task of inferring corresponding initial conditions at an earlier epoch from present cosmological observations. 
This results in a highly non-trivial Bayesian inverse problem, requiring to explore the very high-dimensional and non-linear space of possible solutions to the initial conditions problem from incomplete observations. These parameter spaces typically consist in $10^6$ to $10^7$ parameters, corresponding to the discretized volume elements of the observed domain.

As for \ares{}, the \borg{} algorithm is assuming a specific data model to interpret the galaxy redshift catalogue and infer the three dimensional density field. We do not describe here the full problem solved by \borg{} as such details are already described in \cite{JASCHEBORG2012,JLW15}. We remind here nonetheless the basic assumptions. \borg{} assumes that the distribution of galaxies, after binning in volumetric elements, are Poisson distributed according to some expectation. This expectation, $\lambda_i$, of the galaxy distribution in the voxel $i$ is modelled as
\begin{equation}
	\lambda_i = \bar{N} R_i A (1 + \delta^f_i[\delta^i])^\alpha,
\end{equation}
with $\bar{N}$ the mean galaxy density, $R_i$ the linear response operator including the effects of redshift and target completeness at the voxel $i$, $A$ and $\alpha$ the bias model parameter and $\delta^f_i$ the non-linear density field at the voxel which functionally depends on the initial density field $\delta^i$. The power law bias model is behaving like the linear bias model when $\delta^f_i$ is small compared to one. In this work the relation between $\delta^f$ and $\delta^i$ is given by the 2LPT. As indicated above, in addition to the data model, we put a Gaussian prior on the initial conditions, with a cosmological power spectrum. This Gaussian prior does not enforce Gaussianity of initial conditions. The prior only enforces that without access to data a Gaussian statistics should be followed. But intrinsically non-Gaussian defects in the data would not be erased under this assumption.

Our algorithm explores the posterior distribution of the Fourier modes of $\delta^i$ and the meta-parameter $\bar{N}$. As pointed out previously, the 2LPT describes the one, two and three-point statistics correctly and represents higher-order statistics very well \citep[see e.g. ][]{MOUTARDE1991,BUCHERT1994,BOUCHET1995,SCOCCIMARRO2000,PTHALOS}. Consequently, the \borg{} algorithm
naturally accounts for features of the cosmic web, such as filaments, that are typically associated to higher-order statistics induced by non-linear gravitational structure formation processes. 
Besides higher-order statistics of the density field, this posterior distribution also accounts for survey geometries, selection effects and noise, inherent to any cosmological observation.
The \borg{} algorithm provides full Bayesian uncertainty quantification by exploring this highly non-Gaussian and non-linear posterior distribution via an efficient Hamiltonian Markov Chain Monte Carlo sampling algorithm \citep[see][for details]{DUANE1987,JASCHEBORG2012}. 
As it incorporates an approximate model of large scale dynamics, it automatically and fully self consistently infers the dynamical evolution of the large scale structure from observations.
In this fashion the algorithm provides dynamical structure formation \textit{histories} compatible with both data and model.
In order to account for luminosity dependent galaxy bias and to make use of automatic noise calibration, we will
further use modifications introduced to the original \borg{} algorithm by \citet{JLW15}.

\section{the Bayesian analysis}
\label{sec:Bayesian_inference}

The analysis of the 2M++ galaxy sample has been performed on a cubic Cartesian domain with a side length of 600\Mpch{} consisting of \(256^3\)  equidistant grid nodes, resulting in \(\sim 1.6\times 10^7\) inference parameters for both the \ares{} and the \borg{} runs. Thus the inference procedure provides data constrained realizations for final (and the initial density fields in the case of \borg{}) at a grid resolution of about \(\sim 2.3\) \Mpch. To integrate the effect of the growth of large scale structure and the cosmological Doppler effects, we assume a fixed standard \(\Lambda\)CDM cosmology with the following set of cosmological parameters (\(\Omega_m=0.3175\), \(\Omega_{\Lambda}=0.6825\), \(\Omega_{b}=0.049\), \(h=0.6711\), \(\sigma_8=0.8344\), \(n_s=0.9624\)) taken from \citet{PLANCK2013_16}. Additionally, for the \borg{} runs, cosmological power-spectra for initial density fields were calculated following the prescription provided by \cite{EH98} and \cite{EH99}. For the \ares{} runs the cosmological power spectrum, the bias values and the mean densities have been left free.
Also note that to guarantee a sufficient resolution of the final density field, we oversample the initial density field by a factor of eight, which requires to evaluate the 2LPT model with \(512^3\) particles.
The algorithm correctly accounts for the displacement of matter in the course of structure formation by inferring initial density fields at their Lagrangian coordinates, while final density fields are recovered at corresponding final Eulerian coordinates. We note that redshift space distortions {are not modelled in the \borg{} algorithm and thus are not accounted for explicitly.} In its present formulation the \borg{} algorithm interprets features associated to redshift distortions as noise
and will tend to infer isotropic density fields. Isotropy of density fields is naturally imposed by assuming diagonal covariance matrices for initial density fields. 
{Adding the treatment of redshift distortions, both small scale and large scale, is not trivial. The redshift distortions on large scale induces a change in the likelihood where the initial conditions appears twice (in the density field and the way it is evaluated). An illustration of the expected important of such effect is given and discussed in Section~\ref{sec:inf_density_field}. The distortions on small scales, dubbed "finger-of-god" \citep[first observationally noted by][]{JacksonFOG}, are even more complicated to model, and causes spreading of the mass of haloes on a large volume. This effect not only depends on scale but also depends on the density regime under consideration. As demonstrated by \citet{Leclercq2014A} cosmic voids reconstructed by the \borg{} algorithm do not show any sign of redshift space distortions. With regard to reconstructed haloes tests on $N$-body simulations showed a remaining residual of 15 percent redshift space distortions at the high mass end.}
In total we generated 6552 samples data constrained realizations for initial and final density fields.
Generally, the computational costs to generate a single Markov sample are equivalent to about two hundred 2LPT model evaluations. We measured the typical time to produce a single sample to be about 1500 seconds on a Intel Xeon E5-4640 using \(16\) cores.

\begin{table*}
	\begin{center}
    \begin{tabular}{rlccc}
    \hline
      \multicolumn{2}{c}{Sample selection}  & Identifier & $\alpha^\ell$ & $\widetilde{N}^\ell$  \\
      \hline
      \multirow{3}{*}{$12.5 < \ktmpp \leq 12.5$ and} &
        $-25.00< \Mktmpp <-23.67$ & 0 & 1.74 & $(1.04 \pm 0.01) \times 10^{-2}$\\[1pt]
      & $-23.67< \Mktmpp < -22.33$ & 1 & 1.21 & $(8.6 \pm 0.1) \times 10^{-2}$\\[1pt]
      & $-22.33< \Mktmpp <-21.00$ & 2 & 1.00& $(1.37 \pm 0.03) \times 10^{-1}$ \\[1pt]
      \hline
      \multirow{3}{*}{$\ktmpp \leq 11.5$ and}
      & $-25.00< \Mktmpp <-23.67$ & 3 & 1.70 & $(1.12 \pm 0.01) \times 10^{-2}$\\[1pt]
      & $-23.67< \Mktmpp <-22.33$  & 4 & 1.20 & $(8.60 \pm 0.07) \times 10^{-2}$\\[1pt]
      & $-22.33 < \Mktmpp <-21.00$  & 5 & 1.15 & $(1.24 \pm 0.02) \times 10^{-1}$\\[1pt]
      \hline
    \end{tabular}
    \end{center}
    \caption{\label{tb:Table_Bias} Bias parameters corresponding to the power-law bias model, as described in the text, for six galaxy sub-samples, subdivided according both to their absolute $\ktmpp$-band magnitudes and their apparent magnitudes. The sub-sample 2 is taken as fiducial with a bias set to one.}
\end{table*}

\section{Inference results}
\label{sec:inference_results}
This section describes inference results obtained using our Bayesian analysis on the 2M++ galaxy compilation. As mentioned in Section~\ref{sec:data}, we cannot run a single code to do the entire analysis. Even though that it is mathematically possible, the time complexity would be too high to obtain results in a timely fashion. So we rely on a splitted analysis, using an approximate statistical model (\ares) to derive some of the meta parameters that will be used in the advanced model (\borg). We first present the relevant results of the analysis using the \ares{} code in Section~\ref{sec:results_ares}. Then we describe the 3d density field obtained by the \borg{} code in Section~\ref{sec:inf_density_field}, along with its convergence properties. In Section~\ref{sec:cosmography}, we present the cosmography of the final density field as inferred by \borg{}. Finally, in Section~\ref{sec:local_void_analysis}, we give a quantitative assessment of the presence of the Local Void behind Milky Way's galactic bulge. 

\subsection{Initialization analysis with ARES}
\label{sec:results_ares}
\begin{figure}
	\includegraphics[width=\hsize]{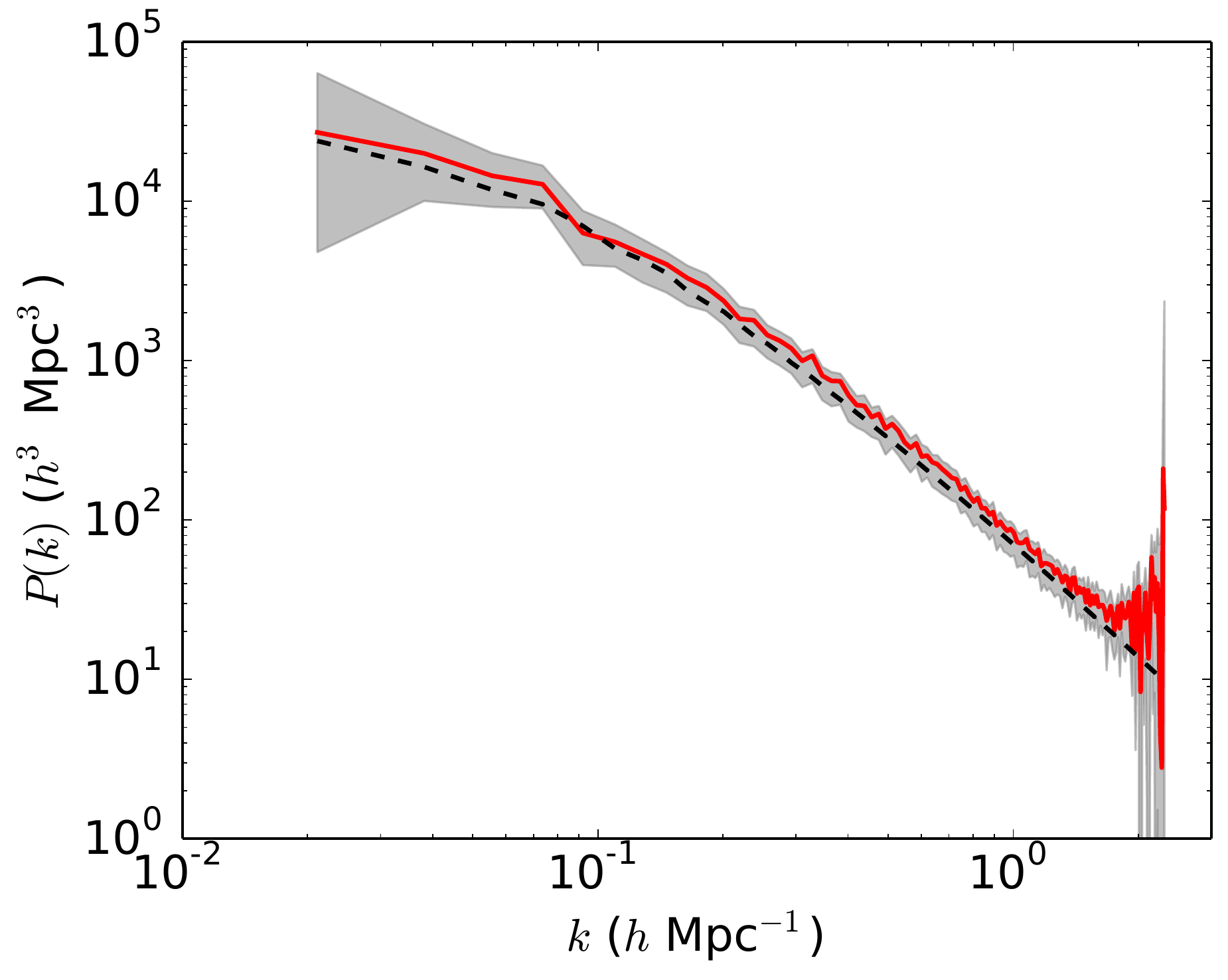}
	\caption{\label{fig:ares_powerspec} Power-spectrum measured by \ares{}. We show here the local of the maximum posterior for each bin in $k$ space (thick red line), alongside the 95\% probability volume (filled grey area) of the power spectrum as measured by \ares{} with the 2M++. The reference power spectrum computed using the \protect\cite{EH98,EH99} approximation including wiggles contributions for the cosmology given in Section~\ref{sec:Bayesian_inference}. }
\end{figure}

As described above, in this work we will use the \ares{} code to calibrate unknown luminosity dependent galaxy biases followed by an detailed analysis with the \borg{} algorithm. To perform this initial analysis with \ares{} we will follow a similar approach as described in \citet{JaschePspec2013}. Specifically we will treat galaxies selected at $\ktmpp \le 11.5$ (sample 1) and $11.5 < \ktmpp \le 12.5$ (sample 2) as two independent data sets with their respective survey geometry and selection function, as detailed in section \ref{sec:data}. In addition we sub divide each of these galaxy samples into three bins of absolute magnitude in the range  $-25.00< \Mktmpp <-21.00$ to account for respective luminosity dependent galaxy biases and noise levels. When applied to the 2M++ data, the \ares{} code generated 4306 joint posterior realizations for the cosmic power-spectrum, the density field, noise levels and luminosity dependent galaxy biases.\footnote{ARES run has been done on a standard workstation Intel Core i7-2600, 8 cores, in a week.} To demonstrate that the \ares{} algorithm inference yielded physically correct results, in Figure \ref{fig:ares_powerspec} we show the comparison between the inferred ensemble mean cosmological power-spectrum and a fiducial one, calculated according to the prescription described in \cite{EH98} and \cite{EH99}. As can be seen \ares{} has recovered the shape of the cosmological power-spectrum within the corresponding one sigma confidence regions. No particular sign of bias throughout all modes in Fourier space can be observed. Erroneous treatment of survey geometries, selection effects and galaxy biases typically yield artefacts of false power in the power-spectrum. The absence of such artefacts in Figure \ref{fig:ares_powerspec}, therefore indicates that these effects have been accounted for accurately. 

\begin{figure}
	\includegraphics[width=\hsize]{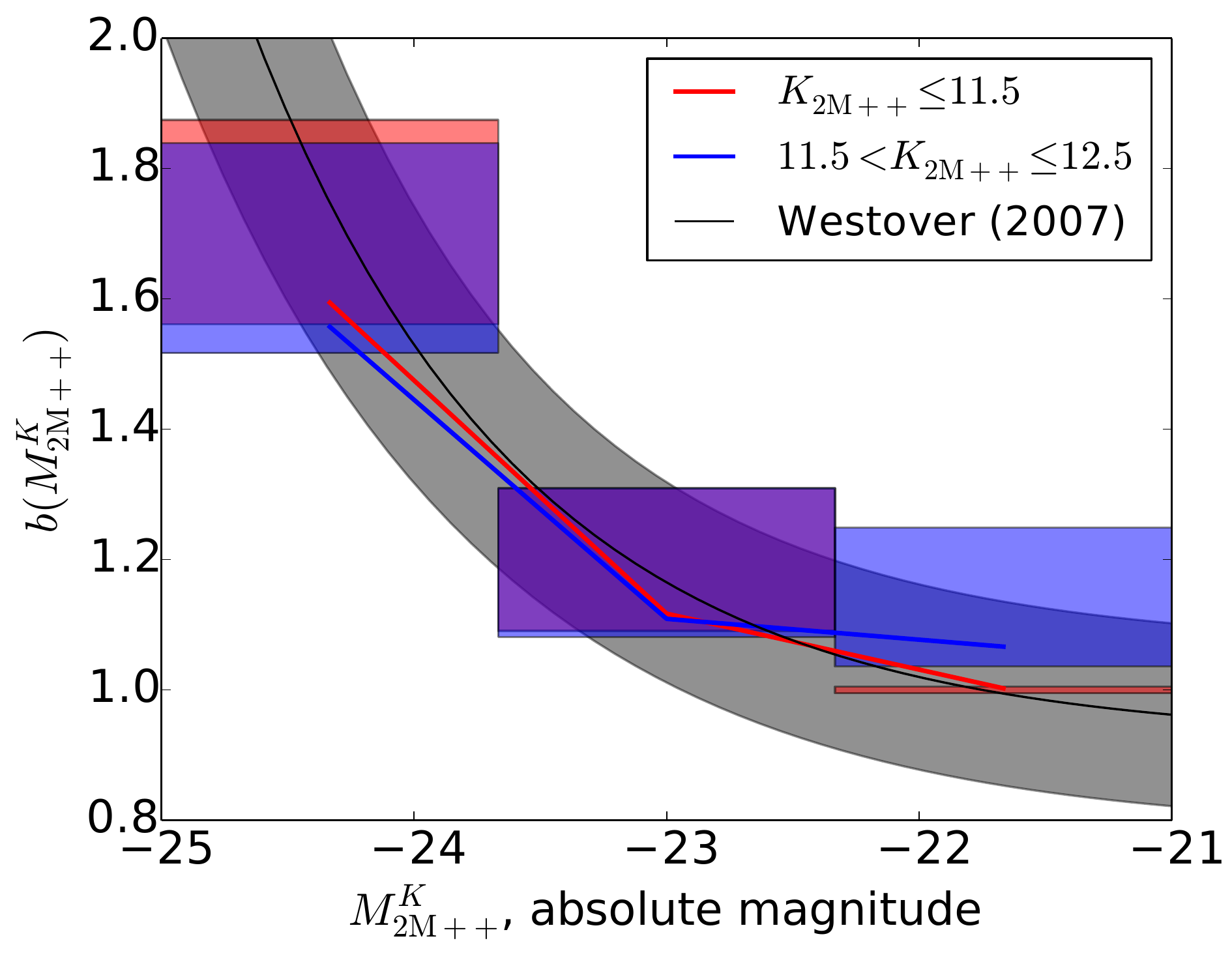}
	\caption{\label{fig:ares_bias} Bias values from \ares{}: We show here the bias values inferred from the 2M++ catalogue assuming a fiducial bias of one for the sub-sample 2 of Table~\ref{tb:Table_Bias} ($11.5 < K_\text{2M++} \leq 12.5$, red).  In addition we have the overplotted the best fit of \protect\cite{Westover07} readjusted for the magnitude bin that serves us as a reference (black). The width of the box gives the interval size of the magnitude bin. Their height gives the 95\% confidence limit of the measurement on bias.}
\end{figure}

In Figure~\ref{fig:ares_bias}, we show the value for the bias parameter found in the different subsample, taking the faintest magnitude bin of the sample 2 with a fiducial value of one. The result are given in red and blue coloured boxes. The width of those boxes corresponds to the width of the magnitude interval and their height to the 95\% confidence interval. In addition, the best fit of \cite{Westover07} have been plotted in black, alongside its error bar analysis. The best fit of \cite{Westover07} is given by 
\begin{equation}
	\frac{b}{b_*} = 0.73 \pm 0.07 + (0.24 \pm 0.04) \frac{L}{L_*}
\end{equation}
with $L$ the intrinsic luminosity of the considered galaxy population, $L_*$ the reference luminosity which for 2M++ is given by $M_* = -23.25$. We have adjusted the reference so that a bias of one is given for our reference population (sample 2, faint luminosity bin). We note the perfect agreement between the two measurement. The advantage of our procedure is its full automation, the derivation of an unbiased power spectrum and the alongside matter density field. Also, we have used a limited number of bins, but nothing prevents us to increase their number, at the cost of the amplitude of the signal-to-noise.
The most important result of the \ares{} analysis for this work is the derivation of the luminosity-dependent galaxy biases for the galaxy population selected in 2M++. {We use these biases as-is in the following \borg{} reconstruction. While the two bias model are relatively different, in the regime of small density fluctuations on large scales, they can be rejoined by doing a Taylor expansion: $(1+\delta_\text{NL})^\alpha \simeq 1 + \alpha \delta_\text{NL}$ and thus $b\simeq \alpha$. Of course this equality is not exact and is probably leading to some bias in the density field reconstruction. We expect in the future to be able to jointly infer the bias parameter in \borg{} with the density field itself at lesser computational cost, which will remove any foreseeable problem.}

\subsection{3d density field}
\label{sec:inf_density_field}

\begin{figure}
\centering{\includegraphics[width=\hsize,clip=true]{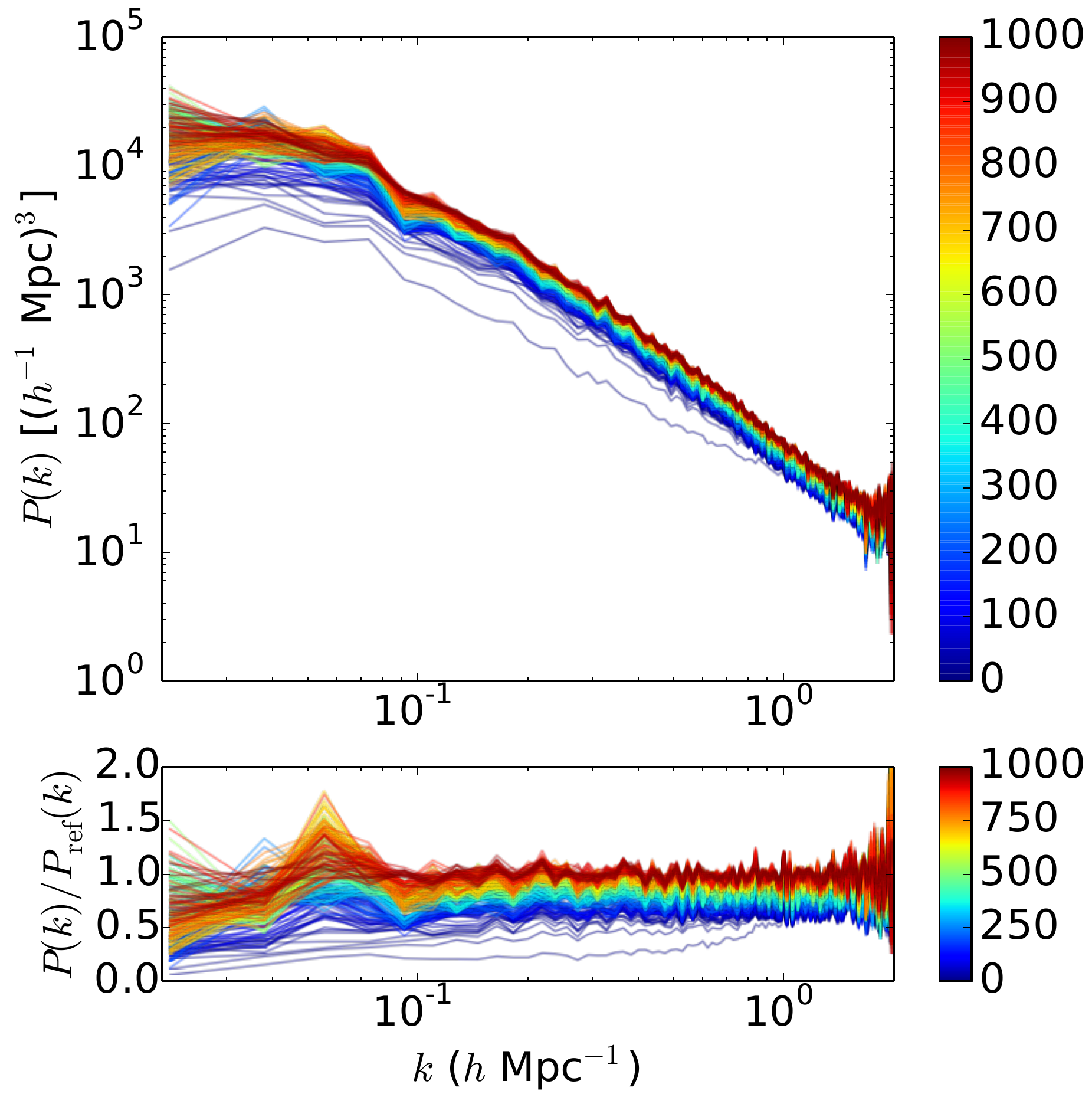}}
\caption{\label{fig:burnin_specs} We are showing here the burn-in phase of the power spectrum. The top panel shows the power spectrum itself, coloured according to the identifier of the step along the Markov Chain. The bottom panel are the same power spectra after having divided {by the assumed \LCDM{} initial linear power-spectrum.} }
\end{figure}

\begin{figure*}
\centering{\includegraphics[width=\hsize,clip=true]{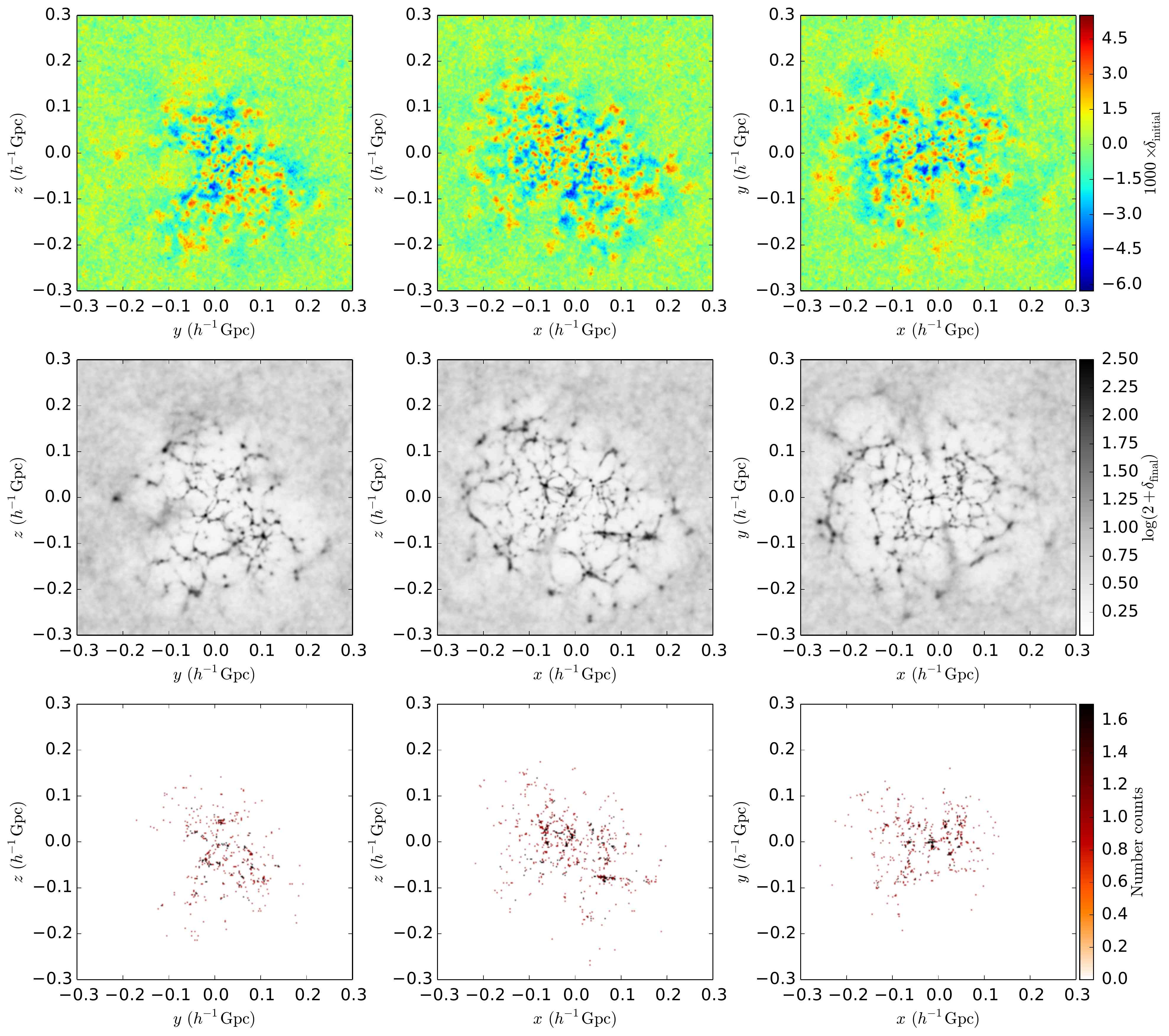}}
\caption{\label{fig:mean_var_dens} Three slices from different directions through the three dimensional ensemble posterior means for the initial (upper panels) and final density fields (middle panels) estimated from 6552 samples. The lower panels depict corresponding slices through the galaxy number counts field of the SDSS main sample. The direction are respectively along the equatorial plane $x=0$ (left-hand column), $y=0$ (middle column) and $z=0$ (right-hand column). }
\end{figure*}

Using inferred bias values, as described above, we have run the \borg{} algorithm on the 2M++ compilation data. The results are presented in Figures~\ref{fig:burnin_specs}, \ref{fig:mean_var_dens} and \ref{fig:supergalactic_plane}.

In Figure~\ref{fig:burnin_specs}, we show the sequence of power-spectra of the initial density field as the chain is attached to a locus around the maximum posterior.  The top panel shows the raw power spectra and the bottom panel are the same power-spectra divided by the assumed \LCDM{} initial linear power-spectrum. We note that after a convergence in $\sim$ 400 samples, the power spectra starts oscillating on large scales ($k \la 0.1$\hMpc). This indicates the chain has extracted all the available information at these scales from observations.  Additionally this indicates the correlation length of the Markov chain to be on the order of $\sim 400$ sampling steps. On intermediate scales ($0.1$\hMpc$ \la k \la 2$\hMpc) the power-spectrum is strongly constrained and unbiased compared to our reference power-spectrum. At very small scale the noise increases back again because we reach scales at most of the size of a voxel element. Consequently all information is lost. We note that, contrary to \cite{Kitaura13}, we do not observe any bumps in the power-spectra of reconstructed phases at intermediate scales.
 Finally we handle unobserved regions sufficiently correctly that the power spectra appear unbiased.

In Figure~\ref{fig:mean_var_dens}, we show the mean initial density field (top row), the 2LPT evolved mean final density field (middle row) and the input data (bottom row) for the $X$, $Y$ and $Z$ plane of the Equatorial coordinate system. The edge of the 2M++ survey is clearly visible in the mean final density field. For these panels, we see clearly defined structures in the central region, which is close to the observer and more likely to be fully complete. Towards the boundaries of the cubic domain structures become increasingly blurry when going out of the observed volume at a distance of $\sim$200\Mpch from the centre. In the initial condition (top row), these edges are far less clear which emphasizes that the information stored in the current position of galaxies comes from extended places in Lagrangian coordinates and that information is distributed differently in initial and final conditions \citep{JLW15}. Finally, we see the visual improvement obtained from the final density field derived by \borg{} compared to the actual distribution of galaxies given in the bottom row. 

\begin{figure*}
    \begin{center}
    	\includegraphics[width=.95\hsize]{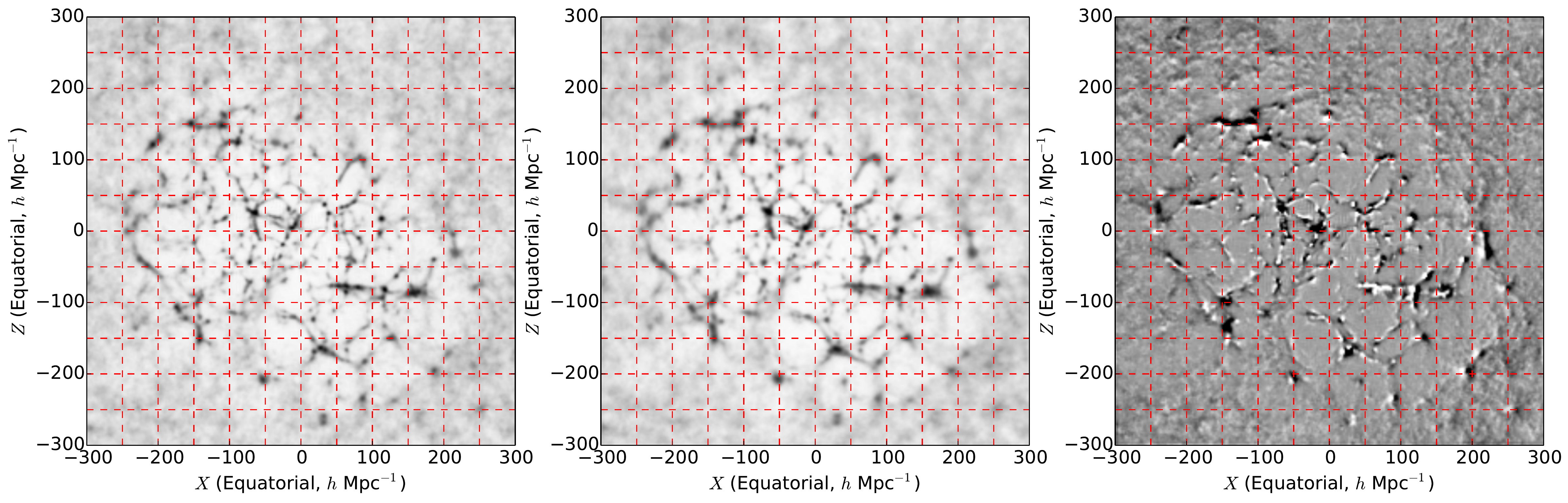}
    \end{center}
	\caption{\label{fig:rsd} {Impact of the large scale component of redshift space distortions on the reconstructed density field. The two panels show the ensemble mean final density fields of the central equatorial slice at $y=0$\Mpch. The left-hand panel gives the real space density field, while the middle panel gives the density field obtained by applying redshift space distortions assuming 2LPT dynamics.  The right-hand panel gives the density difference between the middle panel and the left-hand panel in the same slice. The horizontal and vertical dashed red lines have been drawn to allow for an easier comparison between the two density fields. } }
\end{figure*}

{ In Figure~\ref{fig:rsd}, we show the impact, a posteriori, of the large scale component of redshift space distortions. In particular \emph{for this test we assume} that inferred density fields have been correctly recovered in real-space and add redshift space distortions corresponding to velocities derived through 2LPT dynamics. The left-hand panel of Figure~\ref{fig:rsd} reproduces the real space density field of Figure~\ref{fig:mean_var_dens} (centre column) as determined by \borg{}. The middle panel shows the redshift distortion effects produced by peculiar velocities predicted by the 2LPT dynamics on the density field. The right-hand panel gives the difference between the middle and the left-hand panel, highlighting the regions that have moved due to redshift distortions. On top of the three density fields, we have drawn a red dash-dotted grid with a spacing of 50\Mpch. As can be seen Large Scale Structures are not moved much by the large scale component of the peculiar velocities. The most important effects lead to smearing of filaments and haloes (middle panel), which already happens when comparing 2LPT dynamics to full non-linear solution since 2LPT does not capture shell crossing effects very well. Inspection of the right-hand panel indicates that structures move typically by a few Mpc, which is of the same order as grid resolution used here ($\sim$ 2.3\Mpch). Thus, on scales larger than a single voxel size inferred density fields are not affected much by this effect. We can conclude that for the purpose of density reconstruction that the fields predicted by \borg{} are very close to what they should be if redshift space distortions were taken into account.}

\subsection{Cosmography}
\label{sec:cosmography}

\begin{figure}
	\begin{center}
		\includegraphics[width=.8\hsize]{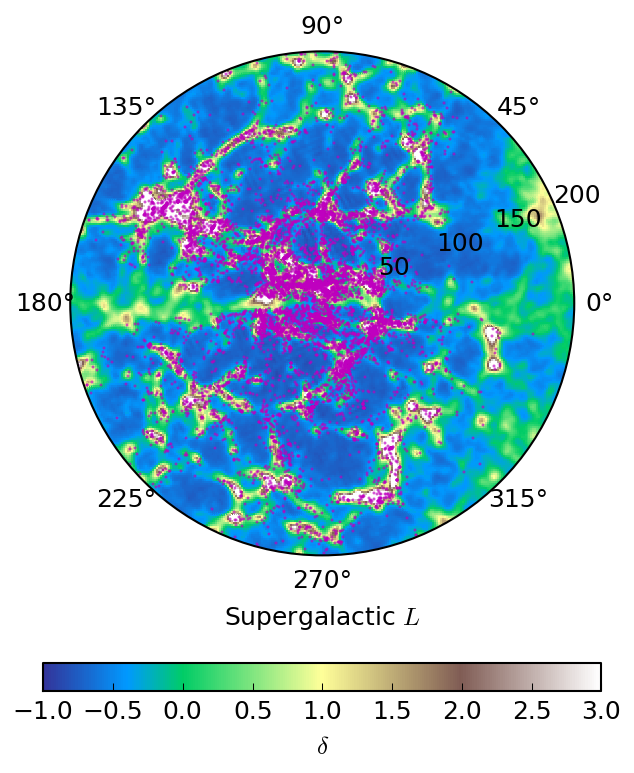}
	\end{center}
	\caption{\label{fig:supergalactic_plane} Supergalactic plane, thickness is 20\Mpch{} for galaxies (shown in magenta points), density is smoothed with a 1\Mpch{} Gaussian kernel (background field with colour scale indicated below the panel).}
\end{figure}

In Figure~\ref{fig:supergalactic_plane}, we show the supergalactic plane as seen from a thin slice of the final density field (coloured background field) computed by \borg{} and a 20\Mpch{}-thick slice (20\Mpch) extracted directly from the galaxy data (magenta dots). We have represented the data in polar coordinates so that the Supergalactic longitude can be directly read from the plot. 

Major structures of the Local Universe are clearly visible both with the galaxies and the final density field. Also, the density field in the Galactic plane (visible at $L=0^\circ$ and $L=180^\circ$) is smoothly extrapolated from neighbouring structures. We typically see the 
Pisces-Cetus supercluster \citep[$L\sim 305\deg$, $d \sim$180\Mpch;][]{Tully86}, the Coma cluster \citep[$L\sim 90\deg$, $d\sim 70$\Mpch;][]{W1901,HH31}, the Shapley concentration  \citep[$L\sim 149\deg$, $d\sim 140$\Mpch;][]{SBCVZ1989,Raychaudhury89} and the Perseus-Pisces supercluster \citep[$L\sim 343$, $d\sim $55\Mpch;][]{JET78}. We note that a quite prominent circular filament connected to the Shapley concentration, going from $L\sim 100\deg$ to $L\sim 150\deg$ at $d\sim  140$\Mpch{}{, located just behind the Bootes void}. We are not aware of any name given to this filament, we name it the Virgo-Bootes-Hercules filament.  

As a final remark, we note that the Sloan Great Wall is clearly visible in the reconstructed density field shown in the middle right-hand panel of Figure~\ref{fig:mean_var_dens}  at $x\sim 225$\Mpch{}, $y\sim 0$\Mpch{}. The wall itself is not clearly visible in the galaxy distribution shown in the panel just below. We see that the Sloan Great Wall is not as well characterized as other structures by looking at the amplitude of the mean field, which is expected given the sparsity of galaxies in the catalogue in that part of the volume. This structure is a striking example of the large-scale structure reconstruction achieved by \borg{} from noisy data. By representing the galaxies and the reconstructed Sloan Great Wall on the same sky plot, we see that the Hercules-Aries filament inters

\subsection{Local Void analysis}
\label{sec:local_void_analysis}

\begin{figure*}
	\begin{center}
		\includegraphics[width=.45\hsize]{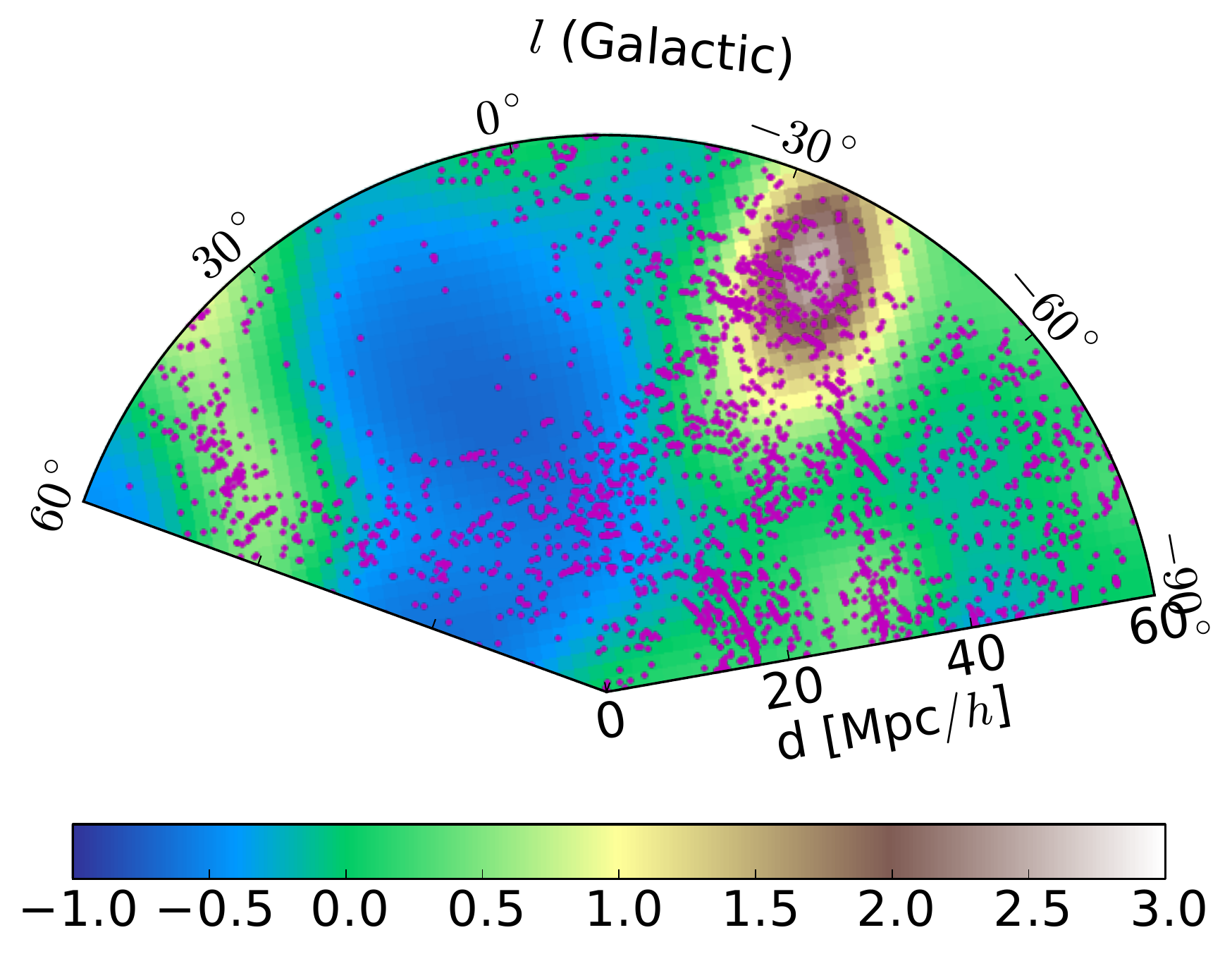}
		\includegraphics[width=.45\hsize]{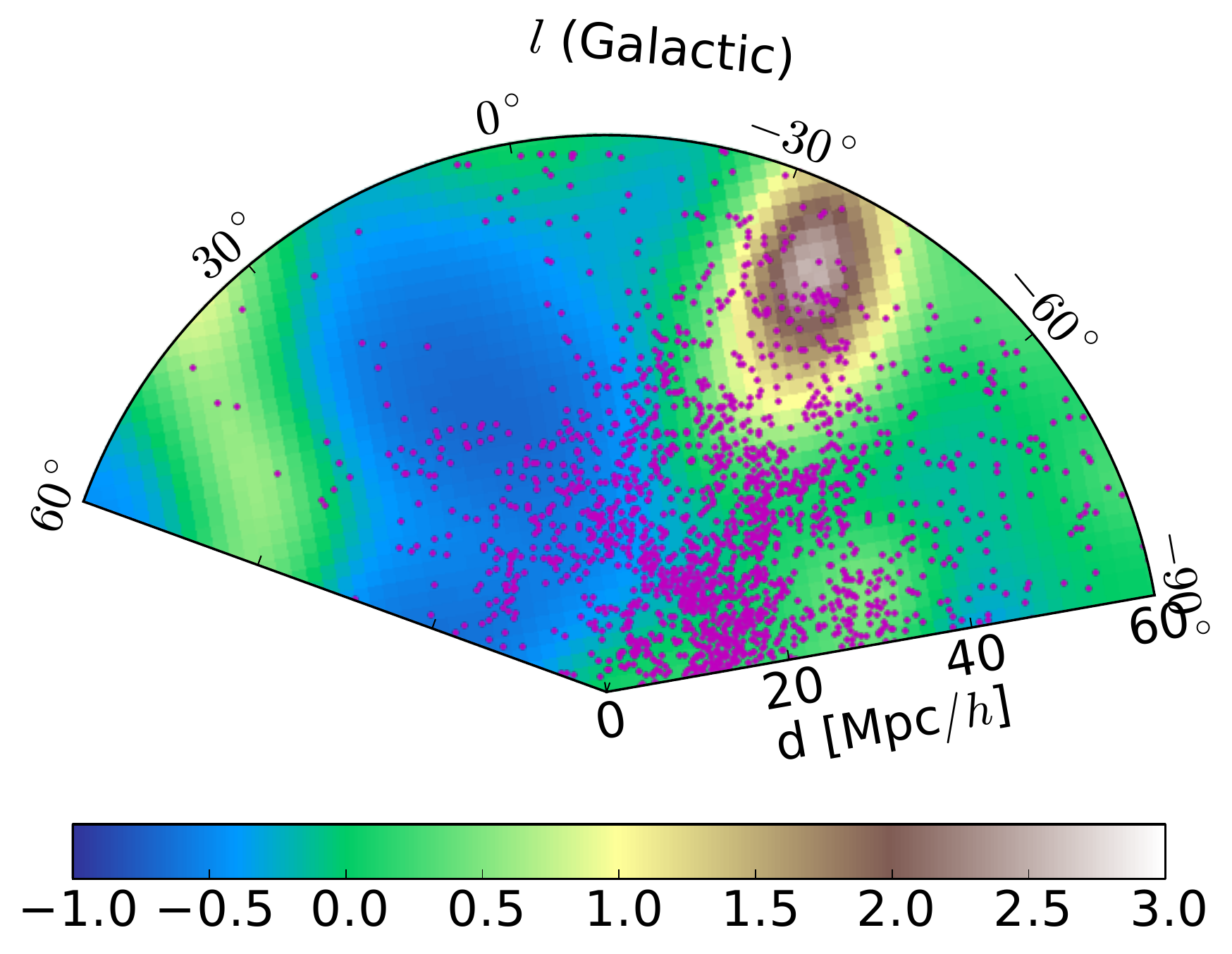}
	\end{center}
	\caption{\label{fig:wedge_galactic_plane} Galactic plane $b=0$, 2M++ left, HICAT right, thickness is 40\Mpch for galaxy representation, the density is smoothed with a 1\Mpch{} Gaussian kernel.}
\end{figure*}

An interesting feature of non-linear density fields inferred by \borg{} is the possibility to uncover unobserved structures. In Figure~\ref{fig:wedge_galactic_plane} we provide a particular example by looking at the Local Void \citep[also known as Tully's void,][]{TF87}. In the two panels we show the mean "final density field" and overplotted by either the 2M++ galaxies (left-hand panel) or the HI Parkes All Sky Survey (HIPASS) galaxies \citep[right-hand panel,][]{HIPASS}. The 2M++ galaxies are appearing in spite of the galactic plane cut and the galactic bulge because we represent a 40\Mpch{} thick slice. This void is clearly visible at the Galactic longitude $l\sim 15^\circ$ in both panels and it visually seems to extend from 10\Mpch{} to 60\Mpch{} in the ensemble mean field. 

To illustrate a further application of our reconstruction technique, we identify and assign a probabilistic value to belonging in a \diva \citep{LW10} void for voxels located in the galactic plane. We have used the following procedure. 

First we smooth the initial density field  of each sample of the Markov Chain created by \borg{} with a Gaussian filter of 5\Mpch{}. The choice of this filter size is motivated by the mass it corresponds to in Lagrangian coordinates. For a Universe with $\Omega_\text{M}=0.30$, a tophat filter of 5\Mpch{} would represent $\sim 4\times 10^{13}\;\text{M}_\odot$. So filtering over that scale removes the contribution from groups of galaxies in the classification of the cosmic web. 

Then we run the truncated watershed transform on this field. We identify particles belonging to the identified voids and propagate forward in time using 2LPT. We set to one each voxel where a void particle is found, and we compute the average field. By construction the average field becomes the marginalized probability for each voxel to be in a void:
\begin{multline}
   \bar{f}^V_p = \frac{1}{C} \sum_{i=1}^{C} f^V_{p}(\{\delta_{q,i}\}) \underset{C\rightarrow +\infty}{=} \int_{\{\delta_q\}} f^V_p(\{\delta_q\}) P(\delta | \text{data}) \text{d}^N \delta_q \\ 
   = P( \text{$p$ is in a void} | \text{data}),
\end{multline}
where $C$ is the length of the Markov Chain, $f^V_{p}(\{\delta_{q}\})$ is set to one if the voxel $p$ belongs to a void assuming initial density fluctuations $\{ \delta_q \}$ and zero otherwise, $P(\delta | \text{data})$ the conditional marginalized posterior of the reconstructed initial density fluctuations given the data. The mean field $\bar{f}^V_p$ is thus equal to the probability that $p$ is in a void given the observational data.

\begin{figure}
	\begin{center}
		\includegraphics[width=.9\hsize]{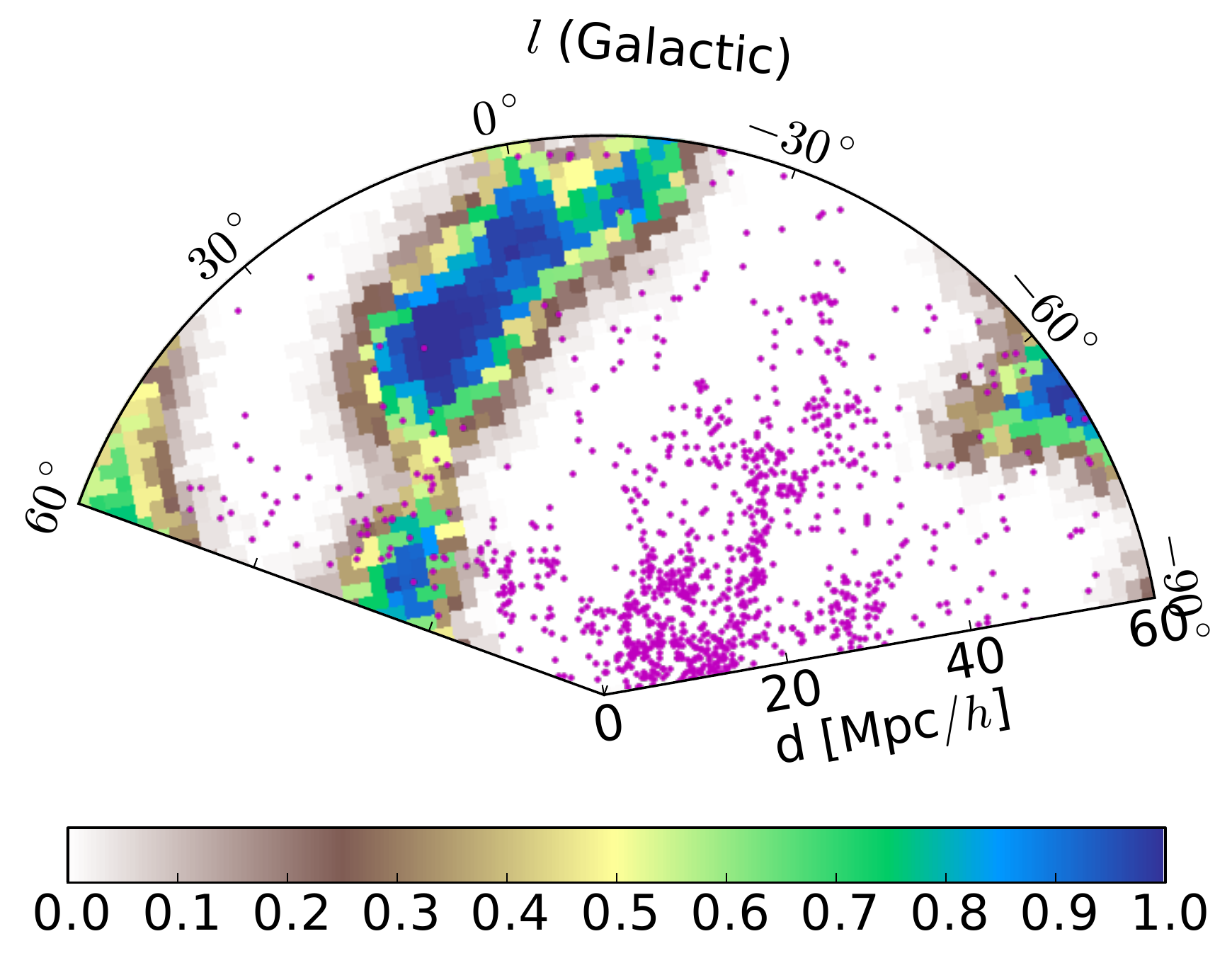}
	\end{center}
	\caption{\label{fig:wedge_void_proba} We show here the probability of a voxel of the galactic plane to belong to a void. The probability goes from "certainly void" (dark blue) to "certainly non-void" (white). We have overplotted the galaxies from the HIPASS survey in magenta shade. The galaxies have been selected such that they are within $10$\Mpch{} from the galactic plane.}
\end{figure}

We show the result of this procedure in the Figure~\ref{fig:wedge_void_proba}, highlighting the regions definitely voids (dark blue colour) or not voids (white). We have over-plotted the galaxies of the HIPASS catalogue that are within 10\Mpch{} of the galactic plane. Of course the regions with a large number of galaxies are more clearly not voids. On the other hand there is a filament of galaxies at $l\sim 60^\circ$ that is marked as belonging to a void with a high probability, i.e. greater than 90\%. We note that the void classification probability is entirely marginalized according to \emph{all} the other variables. The classification here corresponds qualitatively well with the visual impression of Figure~\ref{fig:wedge_galactic_plane} for which the void-like area located in the most under-dense region at longitudes between $\sim 0^\circ$ and $\sim 30^\circ$. Most of the voxels to the right of $0^\circ$ are identified as non-void. 
Of course this classification is not the full story, and it has been advocated by \cite{LW10} that one should use a full filtering hierarchy to characterize dynamically the cosmic web. It is however a powerful tool to separate the galaxies according to their dynamical environment. As it would be beyond the scope of this paper, we postpone this classification to a future work.
We also note that the \diva{} classification of the Large Scale structure is not unique as other prescriptions have been advocated in other work that rely only on the present gravitational field \citep[such as][]{HAHN2007}. However the combination of \borg{} and \diva{} allows us to use the full dynamical history of Large Scale structures to make the classification. Contrary to other techniques, it accounts for the fact that galaxies may have originally formed in environments different from their present one.

\section{Summary and Conclusions}
\label{sec:Summary_an_Conclusion}

This work presents a fully Bayesian data analysis pipeline to study cosmic structures in galaxy redshift catalogues, derive their statistical properties and infer corresponding initial conditions as well as plausible dynamic structure formation histories. This pipeline consists in the sequential application of two of our Bayesian inference algorithms.

Specifically, here we have applied this methodology to the 2M++ galaxy compilation \citep{LH11}, spanning the entire sky at a depth 
of $\sim$ 200\Mpch{}. 
In a first step we have employed the \ares{} \citep{JaschePspec2013} algorithm to infer the cosmological power-spectrum and calibrate luminosity dependent galaxy biases. As demonstrated in Section \ref{sec:results_ares} the \ares{} algorithm accurately recovers the shape of a fiducial cosmological power-spectrum throughout the entire range of Fourier modes considered in this work. This result clearly demonstrates that systematics arising from survey geometries, selection effects and galaxy biases have been accounted for in our Bayesian inference approach. In particular, we have determined the bias values of galaxies with luminosities in three bins for magnitudes going from  $K_\text{2M++}~=~-25.00$ to $K_\text{2M++}~=~-21.00$.
We note that our results on luminosity dependent galaxy biases are consistent with and confirm the previous findings of \cite{Westover07}.

Based upon these results we performed a highly detailed analysis of the mildly non-linear and non-linear large scale structure in the 2M++ galaxy catalogue via the
\borg{} algorithm \citep{JASCHEBORG2012}. 
Specifically, we have used the previously inferred galaxy biases as an input to \borg{} to infer the large scale structure of the Nearby Universe within a co-moving equidistant box of a volume of (600\Mpch)$^{3}$ centred on the observer. The grid resolution is $\sim 2.3$ \Mpch, resulting in a total of $\sim 1.6\times 10^7$ inference parameters which can be accurately handled by our Bayesian inference framework. The algorithm jointly infers the present non-linear Large Scale structures and their corresponding initial conditions, at a cosmic scale factor of $a\sim10^{-3}$, from which they originate. 
In Section \ref{sec:inf_density_field} we have demonstrated the results for inferred three dimensional density fields. These results show highly detailed Large Scale structures at present and in initial conditions. Further we have shown that our Bayesian inference algorithm permits us to accurately quantify uncertainties inherent to any cosmological observations. We have thus successfully reconstructed statistically the initial conditions on large scales of our Local Universe together with a detailed treatment of survey geometries, selection effects and tracer biases. 

As a particular application of the reconstructed density field and initial conditions to statistical structure detection, we have focused on the problem of identifying the Local Void. The Local Void is typically obscured by the Galaxy and is consequently masked out in the 2M++ galaxy compilation. To demonstrate the power of our Bayesian methodology to recover structures in unobserved regions we have shown that the Local Void is clearly visible in the reconstructed density field at $z=0$ despite the lack of information.
To further quantify the statistical significance of this detection, we have used the \diva{} void classification prescription to generate a density of probability  that a given volume element is part of the Local Void. These results indicate a high probability for the existence of the Local Void behind the Galaxy. The validity of our results is further supported by comparison with data from the HIPASS catalogue.  

The results obtained in this work will be subject to more detailed studies, including further improvement in the dynamical model used in the \borg{} tool, of the large scale structure in the Nearby Universe.

In summary, this work presents a detailed application of our  Bayesian inference framework to data of the 2M++ galaxy catalogue. 
In contrast to state-of-the-art approaches, our algorithm accurately recovers 
structures in noisy and masked regimes and also infers the dynamic formation history of individual large scale structures. As a result this methodology opens new windows to analyse and understand the Large Scale structures of our Universe.

\section*{Acknowledgements}
Special thanks go to St\'ephane Rouberol for his support during the course of this work, in particular for guaranteeing flawless use of all required computational resources.
JJ is partially supported by a Feodor Lynen Fellowship by the Alexander von Humboldt foundation and Benjamin Wandelt's Chaire d'Excellence from the Agence Nationale de la Recherche. This research was supported by the DFG cluster of excellence "Origin and Structure of the Universe" (www.universe-cluster.de).
This work made in the ILP LABEX (under reference ANR-10-LABX-63) was supported by French state funds
managed by the ANR within the Investissements d'Avenir programme under reference ANR-11-IDEX-0004-02.
The Parkes telescope is part of the Australia Telescope which is funded by the Commonwealth of Australia for operation as a National Facility managed by CSIRO.
This work was granted access to the HPC resources of The Institute for scientific Computing and Simulation financed by Region \^Ile-de-France and the project Equip\@Meso (reference ANR-10-EQPX-29-01) overseen by the French National Research Agency (ANR) as part of the ``Investissements d'Avenir'' program. We acknowledge financial support from "Programme National de Cosmologie and Galaxies" (PNCG) of CNRS/INSU, France.


\begin{thebibliography}{}

\bibitem[\protect\citeauthoryear{{Abazajian}, {Adelman-McCarthy},
  {Ag{\"u}eros}, {Allam}, {Allende Prieto}, {An}, {Anderson}, {Anderson},
  {Annis}, {Bahcall} \& et al.}{{Abazajian} et~al.}{2009}]{SDSS7}
{Abazajian} K.~N.,  {Adelman-McCarthy} J.~K.,  {Ag{\"u}eros} M.~A.,  {Allam}
  S.~S.,  {Allende Prieto} C.,  {An} D.,  {Anderson} K.~S.~J.,  {Anderson}
  S.~F.,  {Annis} J.,  {Bahcall} N.~A.,    et al. 2009, \apjs, 182, 543

\bibitem[\protect\citeauthoryear{{Ahn}, {Alexandroff}, {Allende Prieto},
  {Anders}, {Anderson}, {Anderton}, {Andrews}, {Aubourg}, {Bailey}, {Bastien}
  \& et al.}{{Ahn} et~al.}{2014}]{SDSS10}
{Ahn} C.~P.,  {Alexandroff} R.,  {Allende Prieto} C.,  {Anders} F.,  {Anderson}
  S.~F.,  {Anderton} T.,  {Andrews} B.~H.,  {Aubourg} {\'E}.,  {Bailey} S.,
  {Bastien} F.~A.,    et al. 2014, \apjs, 211, 17

\bibitem[\protect\citeauthoryear{{Blanton}, {Eisenstein}, {Hogg}, {Schlegel} \&
  {Brinkmann}}{{Blanton} et~al.}{2005}]{BLANTON2005}
{Blanton} M.~R.,  {Eisenstein} D.,  {Hogg} D.~W.,  {Schlegel} D.~J.,
  {Brinkmann} J.,  2005, \apj, 629, 143

\bibitem[\protect\citeauthoryear{{Bouchet}, {Colombi}, {Hivon} \&
  {Juszkiewicz}}{{Bouchet} et~al.}{1995}]{BOUCHET1995}
{Bouchet} F.~R.,  {Colombi} S.,  {Hivon} E.,    {Juszkiewicz} R.,  1995, \aap,
  296, 575

\bibitem[\protect\citeauthoryear{{Buchert}, {Melott} \& {Weiss}}{{Buchert}
  et~al.}{1994}]{BUCHERT1994}
{Buchert} T.,  {Melott} A.~L.,    {Weiss} A.~G.,  1994, \aap, 288, 349

\bibitem[\protect\citeauthoryear{{Duane}, {Kennedy}, {Pendleton} \&
  {Roweth}}{{Duane} et~al.}{1987}]{DUANE1987}
{Duane} S.,  {Kennedy} A.~D.,  {Pendleton} B.~J.,    {Roweth} D.,  1987,
  Physics Letters B, 195, 216

\bibitem[\protect\citeauthoryear{{Eisenstein} \& {Hu}}{{Eisenstein} \&
  {Hu}}{1998}]{EH98}
{Eisenstein} D.~J.,  {Hu} W.,  1998, \apj, 496, 605

\bibitem[\protect\citeauthoryear{{Eisenstein} \& {Hu}}{{Eisenstein} \&
  {Hu}}{1999}]{EH99}
{Eisenstein} D.~J.,  {Hu} W.,  1999, \apj, 511, 5

\bibitem[\protect\citeauthoryear{{Elsner} \& {Wandelt}}{{Elsner} \&
  {Wandelt}}{2013}]{2013A&A...549A.111E}
{Elsner} F.,  {Wandelt} B.~D.,  2013, \aap, 549, A111

\bibitem[\protect\citeauthoryear{{Eriksen}, {O'Dwyer}, {Jewell}, {Wandelt},
  {Larson}, {G{\'o}rski}, {Levin}, {Banday} \& {Lilje}}{{Eriksen}
  et~al.}{2004}]{Eriksen04}
{Eriksen} H.~K.,  {O'Dwyer} I.~J.,  {Jewell} J.~B.,  {Wandelt} B.~D.,  {Larson}
  D.~L.,  {G{\'o}rski} K.~M.,  {Levin} S.,  {Banday} A.~J.,    {Lilje} P.~B.,
  2004, \apjs, 155, 227

\bibitem[\protect\citeauthoryear{{Hahn}, {Porciani}, {Carollo} \&
  {Dekel}}{{Hahn} et~al.}{2007}]{HAHN2007}
{Hahn} O.,  {Porciani} C.,  {Carollo} C.~M.,    {Dekel} A.,  2007, \mnras, 375,
  489

\bibitem[\protect\citeauthoryear{{Hubble} \& {Humason}}{{Hubble} \&
  {Humason}}{1931}]{HH31}
{Hubble} E.,  {Humason} M.~L.,  1931, \apj, 74, 43

\bibitem[\protect\citeauthoryear{{Huchra}, {Macri}, {Masters}, {Jarrett},
  {Berlind}, {Calkins}, {Crook}, {Cutri}, {Erdo{\v g}du}, {Falco}
  et~al.,}{{Huchra} et~al.}{2012}]{Huchra12}
{Huchra} J.~P.,  {Macri} L.~M.,  {Masters} K.~L.,  {Jarrett} T.~H.,  {Berlind}
  P.,  {Calkins} M.,  {Crook} A.~C.,  {Cutri} R.,  {Erdo{\v g}du} P.,  {Falco}
  E.,    et~al., 2012, \apjs, 199, 26

\bibitem[\protect\citeauthoryear{{J{\~o}eveer}, {Einasto} \&
  {Tago}}{{J{\~o}eveer} et~al.}{1978}]{JET78}
{J{\~o}eveer} M.,  {Einasto} J.,    {Tago} E.,  1978, \mnras, 185, 357

\bibitem[\protect\citeauthoryear{{Jackson}}{{Jackson}}{1972}]{JacksonFOG}
{Jackson} J.~C.,  1972, \mnras, 156, 1P

\bibitem[\protect\citeauthoryear{{Jasche} \& {Kitaura}}{{Jasche} \&
  {Kitaura}}{2010}]{JASCHE2010HADESMETHOD}
{Jasche} J.,  {Kitaura} F.~S.,  2010, \mnras, 407, 29

\bibitem[\protect\citeauthoryear{{Jasche}, {Kitaura}, {Li} \&
  {En{\ss}lin}}{{Jasche} et~al.}{2010}]{JASCHE2010HADESDATA}
{Jasche} J.,  {Kitaura} F.~S.,  {Li} C.,    {En{\ss}lin} T.~A.,  2010, \mnras,
  409, 355

\bibitem[\protect\citeauthoryear{{Jasche}, {Kitaura}, {Wandelt} \&
  {En{\ss}lin}}{{Jasche} et~al.}{2010}]{JASCHESPEC2010}
{Jasche} J.,  {Kitaura} F.~S.,  {Wandelt} B.~D.,    {En{\ss}lin} T.~A.,  2010,
  \mnras, 406, 60

\bibitem[\protect\citeauthoryear{{Jasche} \& {Lavaux}}{{Jasche} \&
  {Lavaux}}{2015}]{JL14}
{Jasche} J.,  {Lavaux} G.,  2015, \mnras, 447, 1204

\bibitem[\protect\citeauthoryear{{Jasche}, {Leclercq} \& {Wandelt}}{{Jasche}
  et~al.}{2015}]{JLW15}
{Jasche} J.,  {Leclercq} F.,    {Wandelt} B.~D.,  2015, \jcap, 1, 36

\bibitem[\protect\citeauthoryear{{Jasche} \& {Wandelt}}{{Jasche} \&
  {Wandelt}}{2013a}]{JASCHEBORG2012}
{Jasche} J.,  {Wandelt} B.~D.,  2013a, \mnras, 432, 894

\bibitem[\protect\citeauthoryear{{Jasche} \& {Wandelt}}{{Jasche} \&
  {Wandelt}}{2013b}]{JaschePspec2013}
{Jasche} J.,  {Wandelt} B.~D.,  2013b, \apj, 779, 15

\bibitem[\protect\citeauthoryear{{Jewell}, {Levin} \& {Anderson}}{{Jewell}
  et~al.}{2004}]{Jewell04}
{Jewell} J.,  {Levin} S.,    {Anderson} C.~H.,  2004, \apj, 609, 1

\bibitem[\protect\citeauthoryear{{Jones}, {Read}, {Saunders}, {Colless},
  {Jarrett}, {Parker}, {Fairall}, {Mauch}, {Sadler}, {Watson}, {Burton},
  {Campbell}, {Cass}, {Croom}, {Dawe}, {Fiegert} et~al.,}{{Jones}
  et~al.}{2009}]{Jones09}
{Jones} D.~H.,  {Read} M.~A.,  {Saunders} W.,  {Colless} M.,  {Jarrett} T.,
  {Parker} Q.~A.,  {Fairall} A.~P.,  {Mauch} T.,  {Sadler} E.~M.,  {Watson}
  F.~G.,  {Burton} D.,  {Campbell} L.~A.,  {Cass} P.,  {Croom} S.~M.,  {Dawe}
  J.,  {Fiegert} K.,    et~al., 2009, \mnras, 399, 683

\bibitem[\protect\citeauthoryear{{Kitaura}}{{Kitaura}}{2013}]{Kitaura13}
{Kitaura} F.-S.,  2013, \mnras, 429, L84

\bibitem[\protect\citeauthoryear{{Landy} \& {Szalay}}{{Landy} \&
  {Szalay}}{1993}]{LS93}
{Landy} S.~D.,  {Szalay} A.~S.,  1993, \apj, 412, 64

\bibitem[\protect\citeauthoryear{{Lavaux} \& {Hudson}}{{Lavaux} \&
  {Hudson}}{2011}]{LH11}
{Lavaux} G.,  {Hudson} M.~J.,  2011, \mnras, 416, 2840

\bibitem[\protect\citeauthoryear{{Lavaux} \& {Wandelt}}{{Lavaux} \&
  {Wandelt}}{2010}]{LW10}
{Lavaux} G.,  {Wandelt} B.~D.,  2010, \mnras, 403, 1392

\bibitem[\protect\citeauthoryear{{Leclercq}, {Jasche}, {Sutter}, {Hamaus} \&
  {Wandelt}}{{Leclercq} et~al.}{2015}]{Leclercq2014A}
{Leclercq} F.,  {Jasche} J.,  {Sutter} P.~M.,  {Hamaus} N.,    {Wandelt} B.,
  2015, \jcap, 3, 47

\bibitem[\protect\citeauthoryear{{Meyer}, {Zwaan}, {Webster}, {Staveley-Smith},
  {Ryan-Weber}, {Drinkwater}, {Barnes}, {Howlett}, {Kilborn}, {Stevens},
  {Waugh}, {Pierce}, {Bhathal}, {de Blok}, {Disney}, {Ekers}, {Freeman}
  et~al.,}{{Meyer} et~al.}{2004}]{HIPASS}
{Meyer} M.~J.,  {Zwaan} M.~A.,  {Webster} R.~L.,  {Staveley-Smith} L.,
  {Ryan-Weber} E.,  {Drinkwater} M.~J.,  {Barnes} D.~G.,  {Howlett} M.,
  {Kilborn} V.~A.,  {Stevens} J.,  {Waugh} M.,  {Pierce} M.~J.,  {Bhathal} R.,
  {de Blok} W.~J.~G.,  {Disney} M.~J.,  {Ekers} R.~D.,  {Freeman} K.~C.,
  et~al., 2004, \mnras, 350, 1195

\bibitem[\protect\citeauthoryear{{Moutarde}, {Alimi}, {Bouchet}, {Pellat} \&
  {Ramani}}{{Moutarde} et~al.}{1991}]{MOUTARDE1991}
{Moutarde} F.,  {Alimi} J.,  {Bouchet} F.~R.,  {Pellat} R.,    {Ramani} A.,
  1991, \apj, 382, 377

\bibitem[\protect\citeauthoryear{{Percival}}{{Percival}}{2005}]{PERCIVAL2005}
{Percival} W.~J.,  2005, \mnras, 356, 1168

\bibitem[\protect\citeauthoryear{{Planck Collaboration}}{{Planck
  Collaboration}}{2014}]{PLANCK2013_16}
{Planck Collaboration} 2014, \aap, 571, A16

\bibitem[\protect\citeauthoryear{{Raychaudhury}}{{Raychaudhury}}{1989}]{Raychaudhury89}
{Raychaudhury} S.,  1989, \nat, 342, 251

\bibitem[\protect\citeauthoryear{{Saunders}, {Sutherland}, {Maddox}, {Keeble},
  {Oliver}, {Rowan-Robinson}, {McMahon}, {Efstathiou}, {Tadros}, {White},
  {Frenk}, {Carrami{\~n}ana} \& {Hawkins}}{{Saunders}
  et~al.}{2000}]{Saunders00}
{Saunders} W.,  {Sutherland} W.~J.,  {Maddox} S.~J.,  {Keeble} O.,  {Oliver}
  S.~J.,  {Rowan-Robinson} M.,  {McMahon} R.~G.,  {Efstathiou} G.~P.,  {Tadros}
  H.,  {White} S.~D.~M.,  {Frenk} C.~S.,  {Carrami{\~n}ana} A.,    {Hawkins}
  M.~R.~S.,  2000, \mnras, 317, 55

\bibitem[\protect\citeauthoryear{{Scaramella}, {Baiesi-Pillastrini},
  {Chincarini}, {Vettolani} \& {Zamorani}}{{Scaramella}
  et~al.}{1989}]{SBCVZ1989}
{Scaramella} R.,  {Baiesi-Pillastrini} G.,  {Chincarini} G.,  {Vettolani} G.,
   {Zamorani} G.,  1989, \nat, 338, 562

\bibitem[\protect\citeauthoryear{{Schechter}}{{Schechter}}{1976}]{SCHECHTER1976}
{Schechter} P.,  1976, \apj, 203, 297

\bibitem[\protect\citeauthoryear{{Scoccimarro}}{{Scoccimarro}}{2000}]{SCOCCIMARRO2000}
{Scoccimarro} R.,  2000, \apj, 544, 597

\bibitem[\protect\citeauthoryear{{Scoccimarro} \& {Sheth}}{{Scoccimarro} \&
  {Sheth}}{2002}]{PTHALOS}
{Scoccimarro} R.,  {Sheth} R.~K.,  2002, \mnras, 329, 629

\bibitem[\protect\citeauthoryear{{Skrutskie}, {Cutri}, {Stiening}, {Weinberg},
  {Schneider}, {Carpenter}, {Beichman}, {Capps}, {Chester}, {Elias}, {Huchra},
  {Liebert}, {Lonsdale}, {Monet} et~al.,}{{Skrutskie}
  et~al.}{2006}]{Skrutskie06}
{Skrutskie} M.~F.,  {Cutri} R.~M.,  {Stiening} R.,  {Weinberg} M.~D.,
  {Schneider} S.,  {Carpenter} J.~M.,  {Beichman} C.,  {Capps} R.,  {Chester}
  T.,  {Elias} J.,  {Huchra} J.,  {Liebert} J.,  {Lonsdale} C.,  {Monet} D.~G.,
     et~al., 2006, \aj, 131, 1163

\bibitem[\protect\citeauthoryear{{Tegmark}, {Blanton}, {Strauss}, {Hoyle},
  {Schlegel}, {Scoccimarro}, {Vogeley}, {Weinberg}, {Zehavi}, {Berlind},
  {Budavari}, {Connolly}, {Eisenstein}, {Finkbeiner} et~al.,}{{Tegmark}
  et~al.}{2004}]{TEGMARK_2004}
{Tegmark} M.,  {Blanton} M.~R.,  {Strauss} M.~A.,  {Hoyle} F.,  {Schlegel} D.,
  {Scoccimarro} R.,  {Vogeley} M.~S.,  {Weinberg} D.~H.,  {Zehavi} I.,
  {Berlind} A.,  {Budavari} T.,  {Connolly} A.,  {Eisenstein} D.~J.,
  {Finkbeiner} D.,    et~al., 2004, \apj, 606, 702

\bibitem[\protect\citeauthoryear{{Tully}}{{Tully}}{1986}]{Tully86}
{Tully} R.~B.,  1986, \apj, 303, 25

\bibitem[\protect\citeauthoryear{{Tully} \& {Fisher}}{{Tully} \&
  {Fisher}}{1987}]{TF87}
{Tully} R.~B.,  {Fisher} J.~R.,  1987, {Nearby galaxies Atlas}.
Cambridge University Press

\bibitem[\protect\citeauthoryear{{Wandelt}, {Larson} \&
  {Lakshminarayanan}}{{Wandelt} et~al.}{2004}]{WANDELT2004}
{Wandelt} B.~D.,  {Larson} D.~L.,    {Lakshminarayanan} A.,  2004, \prd, 70,
  083511

\bibitem[\protect\citeauthoryear{Westover}{Westover}{2007}]{Westover07}
Westover M.,  2007, {PhD} dissertation, Harvard University, Department of
  Astronomy

\bibitem[\protect\citeauthoryear{{Wolf}}{{Wolf}}{1901}]{W1901}
{Wolf} M.,  1901, Astronomische Nachrichten, 155, 127

\bibitem[\protect\citeauthoryear{{York}, {Adelman}, {Anderson} Jr., {Anderson},
  {Annis}, {Bahcall}, {Bakken}, {Barkhouser}, {Bastian}, {Berman}, {Boroski},
  {Bracker}, {Briegel}, {Briggs}, {Brinkmann} et~al.,}{{York}
  et~al.}{2000}]{YORK2000}
{York} D.~G.,  {Adelman} J.,  {Anderson} Jr. J.~E.,  {Anderson} S.~F.,  {Annis}
  J.,  {Bahcall} N.~A.,  {Bakken} J.~A.,  {Barkhouser} R.,  {Bastian} S.,
  {Berman} E.,  {Boroski} W.~N.,  {Bracker} S.,  {Briegel} C.,  {Briggs} J.~W.,
   {Brinkmann} J.,    et~al., 2000, \aj, 120, 1579

\end{thebibliography}

\appendix

\bsp
\label{lastpage}

\end{document}